\documentclass[12pt]{iopart}

\usepackage[pdftex]{graphicx}

\def\eqref{\eref}
\newcommand{\mod}{{\rm \ mod\ }}

\usepackage{iopams}
\newcommand{\set}[1]{\left\{ #1 \right\}}
\newcommand{\pdfrac}[2]{\frac{\partial #1}{\partial #2}}
\newcommand{\pddiv}[2]{\partial #1 / \partial #2}
\def\toexp{\mathop{\rm exp}}
\newcommand{\Texp}{\toexp_{\leftarrow}}
\newcommand{\AntiTexp}{\toexp_{\rightarrow}}
\def\Tr{\mathop{\mathrm{Tr}}\nolimits}               
\newcommand{\bra}[1]{\langle{}#1{}|}
\newcommand{\ket}[1]{|{}#1{}\rangle}
\newcommand{\bracket}[2]{\langle{}#1{}|{}#2{}\rangle}

\newcommand{\ketbra}[2]{|{}#1{}\rangle\langle{}#2{}|}

\newcommand{\Cop}{\hat{C}}

\newcommand{\pma}{\mathfrak{S}}
\newcommand{\nuK}{\nu}
\newcommand{\AD}[1]{A^{\mathrm{D} #1}}

\begin{document}

\title[Gauge invariants of eigenspace and eigenvalue anholonomies]{%
  Gauge invariants of eigenspace and eigenvalue anholonomies: 
  Examples in hierarchical quantum circuits}

\author{Atushi Tanaka$^1$, Taksu Cheon$^2$, Sang Wook Kim$^3$}
\address{$^1$ Department of Physics, Tokyo Metropolitan University,
  Hachioji, Tokyo 192-0397, Japan}
\ead{tanaka-atushi@tmu.ac.jp}
\address{$^2$ Laboratory of Physics, Kochi University of Technology,
  Tosa Yamada, Kochi 782-8502, Japan}
\address{$^3$ Department of Physics Education, Pusan National University,
  Busan 609-735, South Korea}


\begin{abstract}
A set of gauge invariants are identified for the gauge theory of
quantum anholonomies, which comprise both the Berry phase and an
exotic anholonomy in eigenspaces. We examine these invariants
for hierarchical families of quantum circuits whose qubit size can 
be arbitrarily large. It is also found that a hierarchical
family of quantum circuits generally involves an NP-complete problem.
\end{abstract}


\maketitle

\section{Introduction}
\label{sec:Introduction}
An adiabatic time evolution of a quantum system along a closed loop
may induce a nontrivial change. The best known among them is the phase
anholonomy, which is also referred to as the Berry phase: a stationary
state acquires an extra phase of geometrical origin as a result of an
adiabatic cycle~\cite{Berry-PRSLA-392-45}.  It has been recognized
that the eigenvalue and eigenspace of a bound state can be also
changed by an adiabatic cycle~\cite{Cheon-PLA-248-285}: When we keep
track of an eigenvalue along a cycle, discrepancies may appear between
the initial and finial eigenvalues.  Such a change, which is called an
eigenvalue anholonomy, implies the interchange of the eigenvalues,
which is in fact allowed because of the periodicity of the spectrum
for the cycle. Furthermore, due to the correspondence between
eigenvalues and eigenspaces, the eigenvalue anholonomy implies the
eigenspace anholonomy, where an adiabatic cycle delivers one
stationary state into another stationary state.

The earliest example of the eigenvalue and eigenspace anholonomies
is found in a family of Hamiltonians with generalized point 
interactions~\cite{generalizedPointInteraction}.
Later, it is recognized that there
is another examples of anholonomy in Hamiltonian systems
that 
involve crossings of 
eigenenergies~\cite{Cheon-PLA-374-144,Tanaka-PRA-82-022104}.
Further examples are found recently
in periodically driven systems 
and quantum circuits, where a stationary state is an eigenvector of 
a unitary operator that 
describes
the
evolution for a unit 
time interval%
~\cite{Tanaka-PRL-98-160407,Miyamoto-PRA-76-042115,Tanaka-EPL-96-10005}.
The spectral degeneracies are known to make both the phase
anholonomy~\cite{Wilczek-PRL-52-2111} 
and the eigenspace anholonomies
more interesting.  However, throughout this paper,
we will focus on the case where no spectral degeneracy exists 
along adiabatic cycles.

We have recently developed a unified gauge theoretic formalism of
eigenspace anholonomy~\cite{Cheon-EPL-85-20001}, extending the
Fujikawa's approach to the phase
anholonomy~\cite{Fujikawa-PRD-72-025009}.  The key quantity for the
formalism is the holonomy matrix $M(C)$, whose elements are given by
overlapping integrals between stationary states and their adiabatic
transports along a cyclic path $C$.  Because $M(C)$ is gauge 
{\it co}variant ~\cite{Cheon-EPL-85-20001}, only a part of $M(C)$ is
observable~\cite{Bohm-GPQS-2003}.

In this paper, we 
show how we extract gauge 
{\it in}variants
from $M(C)$.
Let us suppose that all stationary states in $N$-dimensional
Hilbert space are connected by the $N$-repetition of the adiabatic cycle.
It can be shown that $M(C)$, under an appropriate choice of the gauge,
is given in terms of 
two gauge invariants $\pma(C)$ and $\gamma(C)$:
\begin{eqnarray}
  \label{eq:Mdecomposed}
  M(C) = \pma(C)\, \rme^{\rmi{}\gamma(C)/N}
  ,
\end{eqnarray}
where $\pma(C)$ is a permutation matrix that
describes the interchange of 
eigenspaces induced by
the adiabatic
cycle $C$, and $\gamma(C)$ is the Berry phase gained through
the $N$-repetition of the cycle $C$. In general, $M(C)$ is decomposed
into the blocks, each of which
consists of a permutation matrix 
and a geometric phase factor
as shown
in \eqref{eq:Mdecomposed}.

Also, in this paper, we provide examples
of systems 
that exhibit various 
types of
$\pma(C)$
by extending a recursive construction of hierarchical 
quantum circuits~\cite{Tanaka-EPL-96-10005}.
Due to its topological nature, 
$\pma(C)$ remains unchanged under small perturbations%
, and
the eigenspace and eigenvalue anholonomy with various $\pma(C)$ 
are robust against small imperfections 
inevitable in experimental 
implementation.

The plan of this paper is the following. 
We first lay out the concept of the eigenspace and eigenvalue anholonomies 
with the minimal model, a quantum circuit on a qubit, 
in Section~\ref{sec:minimal}.
In Section~\ref{sec:formalism}, we revisit 
our gauge theoretic 
formalism~\cite{Cheon-EPL-85-20001} to reveal 
the hidden gauge invariants $\pma(C)$ and $\gamma(C)$ residing 
in $M(C)$.
In the following sections, we 
attempt to extend 
the recursive 
construction of quantum circuits~\cite{Tanaka-EPL-96-10005} 
in order to obtain 
expanded
instances of families of systems with 
variety of
$\pma(C)$.
In Section~\ref{sec:model}, we introduce 
novel 
families of quantum circuits.
The eigenvalue problem of these circuits are also solved.
In Section~\ref{sec:eigenangle}, we examine the eigenvalue anholonomy 
of our hierarchical models, 
and obtain $\pma(C)$ for each 
family of quantum circuits.
In Section~\ref{sec:eigenspace}, we examine the eigenspace anholonomy 
of the hierarchical models to obtain $\gamma(C)$ hidden in $M(C)$.
In Section~\ref{sec:examples}, 
several examples are shown. 
We 
examine
the relationship 
between
our model and the adiabatic quantum 
computation~\cite{Farhi-quant-ph-0001106} along the eigenangle
\cite{Tanaka-PRA-81-022320}
in Section~\ref{sec:subset-sum}.
In Section~\ref{sec:discussion}, we discuss the relationship 
between our result and
a topological characterization of the winding of quasienergies
for periodically driven systems~\cite{Kitagawa-PRB-82-235114}.
A summary of this paper is 
found
in Section~\ref{sec:summary}.
Several technical details are provided in an Appendix.

\section{Anholonomies in quantum circuits}
\label{sec:minimal}

We lay out the eigenvalue and eigenspace anholonomies and associated
gauge invariants using a family of quantum circuits on a qubit.  These
quantum circuits provide the simplest case for our analysis shown in
latter sections.

We introduce a quantum circuit on 
a qubit, or, equivalently a quantum map for 
spin-$\frac{1}{2}$~\cite{Cheon-EPL-85-20001,TANAKA-AP-85-1340}:
\begin{eqnarray}
  \label{eq:defU}
  \hat{u}(\lambda, p)
  \equiv
  \exp\left\{\rmi{}(p-1)\lambda(1-\ketbra{y}{y})
    +\rmi{}\lambda\ketbra{y}{y}\right\}\hat{Z}
  ,
\end{eqnarray}
where $p$ is an integer. 
Using orthonormal vectors $\ket{0}$ and $\ket{1}$
of the qubit,
we define $\ket{y} \equiv (\ket{0}-i\ket{1})/{\sqrt{2}}$ and
$\hat{Z}\equiv\ketbra{0}{0} - \ketbra{1}{1}$.
Because $\hat{u}(\lambda, p)$ is $2\pi$-periodic in $\lambda$,
we will examine the eigenvalue and eigenspace anholonomies for 
the cycle $C$, 
for which
$\lambda$ is increased from $0$ to $2\pi$.
We denote an eigenvalue of the unitary operator $\hat{u}(\lambda, p)$ as
$\exp\{i\theta(n; \lambda, p)\}$,
where an eigenangle $\theta(n; \lambda, p)$ satisfies
\begin{eqnarray}
  \label{eq:Eigenangle1}
  \theta(n;\lambda, p) 
  = n\pi + \frac{p}{2}\lambda
  ,
\end{eqnarray}
for $n=0, 1$.
The corresponding eigenvectors are the following:
\numparts
\begin{eqnarray}
  \label{eq:EigenvectorOfUp}
  \ket{0(\lambda, p)}
  &
  = \ket{0}\cos\frac{(2-p)\lambda}{4}+\ket{1}\sin\frac{(2-p)\lambda}{4}
  ,\\
  \label{eq:EigenvectorOfDown}
  \ket{1(\lambda, p)}
  &
  =\ket{1}\cos\frac{(2-p)\lambda}{4}-\ket{0}\sin\frac{(2-p)\lambda}{4}
  .
\end{eqnarray}
\endnumparts

First, we examine an anholonomy in the eigenangle 
$\theta(n;\lambda, p)$.
As $\lambda$ is 
increased from $0$ to $2\pi$ along the closed cycle $C$, we obtain the 
following change
\begin{eqnarray}
  \theta(n;\lambda+2\pi, p)
  = \theta(s(n;C);\lambda, p) + 2\pi r(n; C)
  ,
\end{eqnarray}
where we introduce integers
\begin{eqnarray}
  \label{eq:sr_single}
  s(n; C)
  \equiv (n + p) {\mod 2}
  ,
  \quad
  r(n; C)
  \equiv \left[({n+p})/{2}\right]
  ,
\end{eqnarray}
and $[x]$ is the maximum integer less than $x$. Hence, when
$p$ is even, $n$-th eigenangle is transported to itself by 
the cycle $C$. On the other hand, odd $p$ implies the presence
of an anholonomy in the eigenangles, i.e.,
$\theta(n;\lambda+2\pi, p) = \theta(\bar{n};\lambda, p) +2\pi n$,
where $\bar{0}=1$ and $\bar{1}=0$.

The eigenangle anholonomy implies an anholonomy in eigenspaces.
Namely, the adiabatic transport of an eigenspace along the cycle $C$ 
induces the change in eigenvectors. 
Suppose that each eigenvector $\ket{n(\lambda,p)}$ satisfies the parallel 
transport condition within the eigenspace%
~\cite{Stone-PRSLA-351-141},
i.e., 
\begin{equation}
  \label{eq:ptc}
  \bra{n(\lambda,p)}\left[\partial_{\lambda}\ket{n(\lambda,p)}\right]=0 
  .
\end{equation}
This 
condition is, in fact,
satisfied by 
the eigenvectors
~\eqref{eq:EigenvectorOfUp} and~\eqref{eq:EigenvectorOfDown}.
The adiabatic cycle $C$ accordingly transports $\ket{n(0, p)}$ to 
$\ket{n(2\pi, p)}$. 
Because of the correspondence between the eigenangle and the eigenspace,
$\ket{n(2\pi, p)}$ must 
agree with $\ket{s(n; C)(0, p)}$ up to a phase factor.
When $p$ is odd, $\ket{n(2\pi, p)}$ is orthogonal to $\ket{n(0, p)}$ because
of $s(n)\ne n$.
Note that the change of the eigenvector occurs in spite of the
absence of the spectral degeneracy along the path $C$. Hence this
anholonomy is different from  Wilczek-Zee's phase anholonomy. 
Instead, an anholonomy appears in the eigenspace.

The change in the eigenvectors is characterized by the overlapping integrals 
between the initial and the finial eigenvector
\begin{eqnarray}
  \label{eq:MbyOverlap}
  \{M(C)\}_{n',n} \equiv \bracket{n'(0, p)}{n(2\pi, p)} 
  ,
\end{eqnarray}
which we call a holonomy matrix. This 
involves the interchange of 
eigenspaces
and the 
accumulation
of phases.
Although we have assumed the parallel transport condition
for each eigenspace in \eqref{eq:MbyOverlap},
$M(C)$ is gauge covariant with respect to the gauge 
transformation for $\ket{n(\lambda, p)}$, 
as will be explained in Section~\ref{sec:formalism}~\cite{Cheon-EPL-85-20001}.
Due to the gauge covariance, only a part of $M(C)$ is 
observable~\cite{Bohm-GPQS-2003}.

We now look for the gauge invariants, i.e. observables,
of the quantum anholonomies for the adiabatic cycle $C$.
They can be identified easily by investigating $M(C)$ closely.
One is the permutation matrix $\pma(C)$:
\begin{eqnarray}
  \label{eq:pmaBy_s}
  \{\pma(C)\}_{n',n}\equiv \delta_{n',s(n; C)}
  ,
\end{eqnarray}
which reflects 
the interchange of eigenvalues and eigenspaces. 
Other
gauge invariants are 
the
geometric phases. When $M(C)$ is 
diagonal, i.e., the eigenspace and eigenvalue anholonomies are absent, 
the $n$-th diagonal element of $M(C)$ is the Berry phase for the $n$-th
eigenspace. On the other hand, if the eigenspace and eigenvalue 
anholonomies occur, $M(C)$ contains the off-diagonal geometric phases, 
which have been introduced to quantify the phase anholonomy for a
noncyclic-path~\cite{Samuel-PRL-60-2339} under the presence 
of the eigenspace interchange~\cite{Manini-PRL-85-3067}.

Let us look at
the holonomy matrix and the gauge invariants in 
our model~\eqref{eq:defU}.
Since eigenvectors \eqref{eq:EigenvectorOfUp} and \eqref{eq:EigenvectorOfDown} 
satisfy the parallel transport condition~\eqref{eq:ptc},
we obtain
\begin{eqnarray}
  M(C)
  = 
  1
  \cos\frac{\pi(2-p)}{2}
  -iY
  \sin\frac{\pi(2-p)}{2}
  ,
\end{eqnarray}
where $1$ is the identity matrix (which will be omitted below) 
and 
\begin{eqnarray}
   Y 
   = 
   \left(\begin{array}{cc}
     0&-i\\ i&0
   \end{array}\right)
   .
\end{eqnarray} 

Because the integer $p$ is a crucial parameter for the anholonomies
as shown above, it is better to classify the cases 
depending on whether
$p$ is even 
and odd to 
investigate
the gauge invariants.
When $p$ is even, we obtain $s(n; C)=n$ for 
$n=0$ or $1$.
Namely the anholonomies apparently disappear, i.e.,
\begin{eqnarray}
  \label{eq:pma2even}
  \pma(C) 
  = 
  \left(\begin{array}{cc}
    1&0\\0&1
  \end{array}\right)
  .
\end{eqnarray}
The
permutation 
described by
$\pma(C)$ consists
of two disconnected cycles each of which has a unit length.
The holonomy matrix takes block-diagonal form, whose diagonal
blocks are two $1\times 1$ matrices, i.e., 
\begin{eqnarray}
  \label{eq:M_for_even_p}
  M(C)
  = 
  \left(\begin{array}{cc}
    \sigma(0;C)&0\\0&\sigma(1;C)
  \end{array}\right)
  ,
\end{eqnarray}
where $\sigma(n;C)\equiv (-1)^{(2-p)/2}$ 
is the Berry phase
factor
acquired through the adiabatic cycle $C$
for the $n$-th eigenspace.

On the other hand, 
odd $p$ implies  $s(0; C)=1$ and $s(1;C)=0$.
Hence the eigenvalue and eigenspace anholonomies appear explicitly, i.e.,
\begin{eqnarray}
  \label{eq:pma2odd}
  \pma(C) 
  = 
  \left(\begin{array}{cc}
    0&1\\1&0
  \end{array}\right)
  ,
\end{eqnarray}
which
is 
the permutation matrix of the cycle whose length is $2$.
The holonomy matrix can be regarded as a product of
$\pma(C)$ and a diagonal unitary matrix, i.e.,
\begin{eqnarray}
  \label{eq:M_for_odd_p}
  M(C)
  = 
  \pma(C) 
  \left(\begin{array}{cc}
    \sigma(0;C)&0\\0&\sigma(1;C)
  \end{array}\right)
  ,
\end{eqnarray}
where we set $\sigma(n;C)\equiv (-1)^{n+(1-p)/2}$.
Each $\sigma(n;C)$ 
has no geometric significance since it depends on the choice of the gauge.
We can 
thus
construct a gauge invariant quantity 
$\sigma(0;C)\sigma(1;C) = -1$%
. This is the
Manini-Pistolesi's off-diagonal
geometric phase factor, which is an extension
of Berry's phase factor for the case that the permutation of
eigenvectors
takes place along
an adiabatic 
evolution%
~\cite{Manini-PRL-85-3067}.
Once we slightly change the phase of eigenvectors as
$\ket{\tilde{0}(\lambda, p)}\equiv\rmi\ket{0(\lambda, p)}$
and $\ket{\tilde{1}(\lambda, p)}\equiv \ket{1(\lambda, p)}$,
the
holonomy matrix in the corresponding gauge takes
a canonical form, i.e.,
\begin{eqnarray}
  \tilde{M}(C)
  = 
  \pma(C)\,\rme^{\rmi\gamma(C)/2}
  ,
\end{eqnarray}
where $\gamma(C) = \pi p$ agrees with the Manini-Pistolesi off-diagonal
geometric phase, up to a modulo $2\pi$.

So far, we have examined specific families
of quantum circuits%
, namely $\hat{u}(\lambda,p)$~\eqref{eq:defU} with
$\ket{y}=(\ket{0}-\rmi\ket{1})/{\sqrt{2}}$ and 
$\hat{Z}=\ketbra{0}{0} - \ketbra{1}{1}$.
This analysis can be applied to the whole family of $\hat{u}(\lambda,p)$ 
with arbitrary $\ket{y}$ and $\hat{Z}$.
For
the path $C_p$ of the whole family 
$\hat{u}(\lambda,p)$, $C_p$ still has
topological invariant, namely either \eqref{eq:pma2even} or
\eqref{eq:pma2odd}.
Although we
may deform the closed path $C_p$ by varying $\ket{y}$ and $\hat{Z}$ in
$\hat{u}(\lambda,p)$%
, the 
topological invariant $\pma(C)$ remains
intact
unless
the deformed path encounters a spectral degeneracy. 
Hence, for 
every
path $C$ 
obtained by 
deforming
$C_p$ with even $p$, $\pma(C)$ is the unit matrix. 
With odd $p$,
$\pma(C)$ is 
always
the permutation matrix of the cycle whose length is $2$.
Hence the closed paths are divided into 
two families related 
to
the value of $\pma(C)$, 
i.e. the unit or the permutation matrix 
except the paths
that come across spectral degeneracies.

\section{Gauge invariants of the quantum anholonomy}
\label{sec:formalism}

In this section, we first 
review our unified gauge theoretical treatment of quantum anholonomy 
that 
is applied
both to 
the Berry phase and the eigenspace 
anholonomy~\cite{Cheon-EPL-85-20001,TANAKA-AP-85-1340}.
This offers a gauge covariant expression of the holonomy matrix
$M(C)$ using a non-Abelian gauge connection.
Second, we reveal gauge invariants in $M(C)$ using several choices 
of the gauge. In particular, we will show that Manini-Pistolesi's
off-diagonal geometric phase in $M(C)$ is the Berry phase of
a hypothetical closed path.

Because the eigenspace anholonomy involves the interchange of multiple 
eigenspaces, we need to treat basis vectors of the relevant eigenspaces 
simultaneously.
Our starting point 
is the sequence of ordered basis vectors 
\begin{eqnarray}
  \label{eq:arbitray_f}
  f(\lambda)
  = \left[\ket{0(\lambda)}, \dots, \ket{N-1(\lambda)}\right]
  ,
\end{eqnarray}
where $N$ is the dimension of the Hilbert space of our system
and the basis vectors are assumed to be normalized.
A non-Abelian gauge connection is then given as
\begin{eqnarray}
  A(\lambda)\equiv
  \left[f(\lambda)\right]^{\dagger}\rmi\partial_{\lambda}f(\lambda)
  ,
\end{eqnarray}
where
the $(n',n)$-th element of $A(\lambda)$ is 
$\bra{n'(\lambda)}\rmi\partial_{\lambda}\ket{n(\lambda)}$.
The anti-path-ordered integral of $A(\lambda)$ along $C$, 
\begin{eqnarray}
  \label{eq:WbyA}
  W(C)\equiv \AntiTexp\left(-\rmi\int_C A(\lambda)\rmd\lambda\right)
  ,
\end{eqnarray}
describes the parametric change of $f(\lambda)$ along $C$. 
Note that
$W(C) = \{f(0)\}^{\dagger}f(2\pi)$, where $0$ and
$2\pi$ 
correspond to 
the initial and final points, respectively,
in $C$.
We also introduce a diagonal gauge connection
\begin{eqnarray}
  \{\AD{}(\lambda)\}_{n',n}
  \equiv \delta_{n',n}\{A(\lambda)\}_{n',n}
  ,
\end{eqnarray}
whose $n$-th diagonal element is Mead-Truhlar-Berry's gauge 
connection~\cite{Mead-JCP-70-2284,Berry-PRSLA-392-45}
for $n$-th eigenspace.
If the eigenspace anholonomy is absent, the path-ordered integral of
$\AD{}(\lambda)$ along $C$, 
\begin{eqnarray}
  \label{eq:BbyAD}
  B(C)&\equiv \Texp\left(\rmi\int_C \AD{}(\lambda)\rmd\lambda\right)
  ,
\end{eqnarray}
contains the Berry phases. 

In general, neither $W(C)$ nor $B(C)$
is 
gauge invariant.
Indeed, the gauge transformation 
$\ket{n(\lambda)}\mapsto \rme^{\rmi\phi_n(\lambda)}\ket{n(\lambda)}$,
when the spectral degeneracy is absent,
induces the change
$W(C)\mapsto G(0)^{\dagger}W(C)G(2\pi)$ 
and 
$B(C)\mapsto G(2\pi)^{\dagger}B(C)G(0)$,
where $G(\lambda)$ is the diagonal unitary matrix whose $n$-th diagonal
element is $\rme^{\rmi\phi_n(\lambda)}$~\cite{Cheon-EPL-85-20001}.
This means 
that 
\begin{eqnarray}
  \label{eq:Mfactorized}
  M(C) = W(C) B(C)
\end{eqnarray}
is gauge covariant, i.e., 
$M(C)\mapsto G(0)^{\dagger}M(C)G(0)$.
Another proof of the covariance of $M(C)$ is obtained through 
the investigation
of adiabatic time evolution of the basis 
vectors directly~\cite{Cheon-EPL-85-20001,TANAKA-AP-85-1340}.

The holonomy matrix $M(C)$ is 
also expressed as
a product of a permutation matrix $\pma(C)$
and a diagonal unitary matrix $\sigma(C)$, i.e.,
\begin{eqnarray}
  \label{eq:MbyProducts}
  M(C) = \pma(C) \sigma(C)
  ,
\end{eqnarray}
because $W(C)$ involves the interchange of the eigenspaces
and $B(C)$ is a diagonal unitary matrix. 
A permutation matrix can be decomposed into cycles, 
each of which is a cyclic permutation~\cite{cycle}.
When $\pma(C)$ is decomposed into $J$ cycles $\pma_j(C)$ ($j=1,\dots,J$),
$M(C)$ takes a block diagonal structure where $j$-th diagonal 
component is $\pma_j(C) \sigma_j(C)$ with a diagonal unitary
matrix $\sigma_j(C)$ 
under a suitable choice of a sequence of the basis vectors.
In the following, we will focus on a diagonal block, or equivalently,
we assume $J=1$, i.e., $\pma(C)$~\eqref{eq:MbyProducts} describes
a single cycle.
Accordingly, 
Manini-Pistolesi's off-diagonal geometric phase
factor
is 
\begin{eqnarray}
  \label{eq:Gamma_by_sigma}
  \rme^{\rmi\gamma(C)} = \det\sigma(C)
  .
\end{eqnarray}
Both $\pma(C)$ and $\gamma(C)$ are gauge invariants of the adiabatic
cycle $C$.

With an appropriate gauge of the basis vectors, $M(C)$ takes
a simpler form 
\begin{eqnarray}
  \label{eq:M_cannonical}
  M(C) = \pma(C) \rme^{\rmi\gamma(C)/N}
  ,
\end{eqnarray}
i.e., the off-diagonal geometric phase $\gamma(C)$ is equally assigned
to each eigenspace.
This tells us that the two gauge invariants
$\pma(C)$ and $\gamma(C)$ are at the heart of $M(C)$.
We 
thus
show \eqref{eq:M_cannonical} by using a diagonal unitary matrix
$U_{\mathrm{d}}$ that satisfies
\begin{eqnarray}
  \label{eq:M_cannonical_by_GT}
  U_{\mathrm{d}}^{\dagger}\{\pma(C) \sigma(C)\}U_{\mathrm{d}} 
  = \pma(C) \rme^{\rmi\gamma(C)/N}
  ,
\end{eqnarray}
where the left hand side 
describes
a gauge transformation of
\eqref{eq:MbyProducts} induced by $U_{\mathrm{d}}$.
First, we assume that the basis vectors are arranged so as to
satisfy $\{\pma(C)\}_{n',n}=\delta_{n',n+1 \mod N}$.
This is always possible for a cyclic $\pma(C)$.
Second, we set
\begin{eqnarray}
&&
\{U_{\mathrm{d}}\}_{00}=1,
\nonumber \\
&&
\{U_{\mathrm{d}}\}_{11}
=
\{U_{\mathrm{d}}\}_{00}
\{\sigma(C)\}_{00}\rme^{-\rmi\gamma{(C)}/N},
\nonumber \\
&&
\{U_{\mathrm{d}}\}_{22}
=\{U_{\mathrm{d}}\}_{11}\{\sigma(C)\}_{11}\rme^{-\rmi\gamma{(C)}/N},
\\ \nonumber 
&&
\qquad\qquad
\vdots
\\ \nonumber
&&
\{U_{\mathrm{d}}\}_{\!N\!-\!1,N\!-\!1}
=\{U_{\mathrm{d}}\}_{N\!-\!2,N\!-\!2}
\{\sigma(C)\}_{\!N\!-\!2,N\!-\!2}\rme^{-\rmi\gamma{(C)}/N}.
\end{eqnarray}
Hence
it is straightforward to confirm~\eqref{eq:M_cannonical_by_GT}.

Now
we 
construct a gauge that
precisely assigns $\pma(C)$ and $\sigma(C)$ to
$W(C)$ and $B(C)$, respectively.
Let us suppose that such a gauge is chosen.
Accordingly $W(C)$ describes only the interchange of eigenspaces.
On the other hand, the Manini-Pistolesi off-diagonal phase 
$\gamma(C)$ is encoded into $B(C)$ 
because $\det B(C)$ coincides with 
$\rme^{\rmi\gamma(C)}$. 
This gauge offers a way to associate $\gamma(C)$ with
the gauge connection from \eqref{eq:BbyAD}:
\begin{eqnarray}
  \label{eq:gammaByAnn}
  \gamma(C) = \sum_n \int_C\{A(\lambda)\}_{nn}\rmd\lambda
  .
\end{eqnarray}
Note that the argument above does not 
guarantee that \eqref{eq:gammaByAnn} is gauge invariant
because $\det B(C)$ is not gauge invariant in general.
Nevertheless we
will show that $\gamma(C)$ is indeed gauge invariant 
below.

For this
we examine 
the origin of 
the deviation of $W(C)$ from $\pma(C)$ 
in
an arbitrary gauge.
First, 
let us recall
$\{W(C)\}_{n',n}= \bracket{{n'}(0)}{n(2\pi)}$
~\eqref{eq:WbyA}.
Second, because $C$ is a loop in the parameter space,
$\{\pma(C)\}_{n',n} = |\{W(C)\}_{n',n}|^2$ is either $0$ or $1$.
We introduce $s(n;C)$ so 
that
$\ket{n(2\pi)}$ is parallel to $\ket{s(n)(0)}$, as is done in
the previous section. Hence \eqref{eq:pmaBy_s} is also 
valid
here.
Now we obtain 
\(
\{W(C)\}_{n',n}
= \{\pma(C)\}_{n',n}\bracket{s(n)(0)}{n(2\pi)}
.
\)
Thus we need to impose $\bracket{s(n)(0)}{n(2\pi)}=1$ 
using a suitable choice of the phases of the basis vectors,
to have $W(C)=\pma(C)$. 

We introduce a normalized vector $\ket{\xi(\lambda)}$
to define basis vectors $\ket{\xi_n(\lambda)}$ ($n=0,\dots,N-1$),
which satisfy $W(C)=\pma(C)$. 
First, for
Once we impose smoothness on $\ket{\xi(\lambda)}$, 
$\ket{\xi(\lambda)}$ 
outside 
$0\le\lambda<2\pi$ is determined.
In particular, 
the projection 
$\ketbra{\xi(\lambda)}{\xi(\lambda)}$ is
$2\pi N$-periodic in $\lambda$.
Next, from $\ket{\xi(\lambda)}$ one can construct $\ket{\xi_n(\lambda)}$
in the following way; 
$\ket{\xi_{0}(\lambda)}=\ket{\xi(\lambda)}$,
$\ket{\xi_{s(0)}(\lambda)}=\ket{\xi(\lambda+2\pi)}$,
$\ket{\xi_{s(s(0))}(\lambda)}=\ket{\xi(\lambda+4\pi)}$,
and so on.
At the final step of this construction, we obtain
$\ket{\xi_{n_{\mathrm{F}}}(\lambda)}=\ket{\xi(\lambda+2\pi(N-1))}$,
where $n_{\mathrm{F}}$ is defined as $s(n_{\mathrm{F}})=0$.
This imposes $\bracket{\xi_{s(n)}(0)}{\xi_{n}(2\pi)}=1$ except for 
$n=n_{\mathrm{F}}$. 
The case $n=n_{\mathrm{F}}$ 
is
related 
to
the single-valuedness of $\ket{\xi(\lambda)}$ 
including
the phase factor 
in $C^N$, i.e., $N$-repetitions 
along
the closed cycle $C$. 
Here we simply assume that there is a gauge that satisfies
$\ket{\xi(\lambda)}$ is single-valued in $C^N$.

Now we return to \eqref{eq:gammaByAnn} to obtain an intuitive
interpretation of $\gamma(C)$. In the gauge specified by
the basis vectors $\ket{\xi_n(\lambda)}$, we have
$\{A(\lambda)\}_{nn} 
=\rmi\bra{\xi_n(\lambda)}\partial_{\lambda}\ket{\xi_n(\lambda)}$.
Because \eqref{eq:gammaByAnn} accumulates the contributions
from all $n=0,\dots,N-1$ along $C$, this can be 
cast
into
an integration along the extended cycle $C^N$:
\begin{eqnarray}
  \gamma(C)
  &
  \label{eq:GammaByBarC}
  =\int_{C^N}
  \bra{\xi(\lambda)}[\rmi\partial_{\lambda}
  \ket{\xi(\lambda)}]\rmd\lambda
  .
\end{eqnarray}
Now it is straightforward to see $\gamma(C)$ is the Berry phase
of the adiabatic basis vector $\ket{\xi(\lambda)}$ along $C^N$.
Hence it is independent of 
the gauge of $\ket{\xi(\lambda)}$,
despite our specific choice.

\section{A recursive construction 
  of $N$-qubit examples}
\label{sec:model}

In the following sections, we extend the quantum circuit
$\hat{u}(\lambda, p)$~\eqref{eq:defU} to multiple-qubit systems.
Our aim is to 
show
a various realization of
$\pma(C)$ 
in many situations.
In this section, we explain 
how 
to extend it
and solve the 
corresponding
eigenvalue problem.

A crucial ingredient of our extension of $\hat{u}(\lambda,p)$
is the 
so-called
``super-operator'' $\hat{D}_p[\cdot]$ labeled by 
an integer $p$:
\begin{eqnarray}
  \hat{D}_{p}[\hat{U}]\equiv \Cop^y_p[\hat{U}] (\hat{Z}\otimes\hat{1})
  ,
\end{eqnarray}
where
\begin{eqnarray}
  \label{eq:defCop_y}
  \Cop^y_p[U]
  \equiv (\hat{1}-\ketbra{y}{y})\otimes \hat{U}^{p-1} 
  + \ketbra{y}{y}\otimes \hat{U}
\end{eqnarray}
is a generalized 
controlled-$U$      
gate.
Here
the ``axis'' of the 
control-bit is 
given in the
``$y$-direction''.
It is worth noting that one can retrieve $\hat{u}(\lambda,p)$
from $\hat{D}_{p}[\cdot]$ through
\begin{equation}
  \hat{D}_{p}[\hat{U}_{\rm g}(\lambda)]
  = \hat{u}(\lambda, p)\otimes\hat{1}_{\rm A}
  ,
\end{equation}
where 
$\hat{U}_{\rm g}(\lambda)\equiv \rme^{\rmi\lambda}\hat{1}_{\rm A}$
and
$\hat{1}_{\rm A}$ are the global phase gate and the 
the identity operator of the ancilla, respectively.

\begin{figure}
  \centering
  \includegraphics[width=10cm]{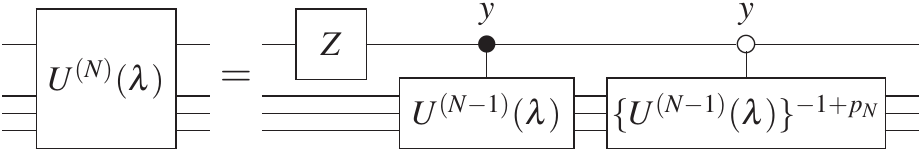}
  \caption{%
    The schematic diagram describing 
    the recursion relation~\eqref{eq:defUN} of $\hat{U}^{(N)}(\lambda)$. 
    The uppermost line indicates the $N$-th qubit, which 
    plays a role of the control qubit of the controlled unitary operations.
    When the control qubit is one (zero)
    the unitary operation connected to 
    the filled (open) circle is applied to the rest of the qubits.
    $y$ above the circles indicates that the axis of the
    control-bit is given in the $y$-direction (see ~\eqref{eq:defCop_y},
    the definition of $\Cop^y_p[U]$).
    In the following, $y$ in circuit diagrams is omitted.
  }
  \label{fig:recursion}
\end{figure}

We
examine a family of quantum circuits 
$\hat{U}^{(N)}(\lambda)$
for a given sequence of integers 
$\{p_j\}_{j=1}^{\infty}$.
We define 
$\hat{U}^{(N)}(\lambda)\equiv \hat{u}(\lambda, p_1)$ for $N=1$,
and
\begin{eqnarray}
  \label{eq:defUN}
  \hat{U}^{(N)}(\lambda)
  \equiv \hat{D}_{p_N}[\hat{U}^{(N-1)}(\lambda)]
\end{eqnarray}
for $N>1$
(see Figure~\ref{fig:recursion}).
If there is no ambiguity, we will omit $p_j$'s.

We now solve the eigenvalue problem of $\hat{U}^{(N)}(\lambda)$ recursively.
Let $\ket{n_N, \cdots, n_1; \lambda}$ 
and $\theta^{(N)}(n_N, \cdots, n_1; \lambda)$ 
denote an eigenvector and the corresponding eigenangle
of $\hat{U}^{(N)}(\lambda)$, i.e.,
\begin{equation}
  \hat{U}^{(N)}(\lambda)\ket{n_N, \cdots, n_1; \lambda}
  = \rme^{\rmi\theta^{(N)}(n_N, \cdots, n_1; \lambda)}
  \ket{n_N, \cdots, n_1; \lambda}
  ,
\end{equation}
where $n_j\in\set{0,1}$.
We have already 
solved
the case $N=1$ in 
Section~\ref{sec:minimal}
(see \eqref{eq:Eigenangle1}, 
\eqref{eq:EigenvectorOfUp} and~\eqref{eq:EigenvectorOfDown}):
$\ket{n_1; \lambda} 
= \ket{n_1(\lambda, p_1)}$
and 
$\theta^{(1)}(n_1; \lambda)= \theta({n_1}; \lambda, p_1)$.
For $N>1$, we
obtain 
recursion relations for eigenangles and eigenvectors.
Suppose we have an eigenvector 
$\ket{\Psi^{(N-1)}(\lambda)}$ and 
the corresponding eigenangle
$\Theta^{(N-1)}(\lambda)$ of $\hat{U}^{(N-1)}(\lambda)$.
Recall
that we obtain $\hat{U}^{(N)}(\lambda)$ 
by 
extending
$\hat{U}^{(N-1)}(\lambda)$
so as to incorporate one more qubit into it.
Let $\ket{\phi}$ denote a state vector for the qubit
newly added
to the quantum circuit described by $\hat{U}^{(N-1)}(\lambda)$.
A natural candidate of an eigenvector of $\hat{U}^{(N)}(\lambda)$ is
$\ket{\phi}\otimes\ket{\Psi^{(N-1)}(\lambda)}$ 
because we have
\begin{eqnarray}
  \hat{U}^{(N)}(\lambda)
  \left(\ket{\phi}\otimes\ket{\Psi^{(N-1)}(\lambda)}\right)
  = \left[\hat{u}(\lambda'; p_N)\ket{\phi}\right]
  \otimes\ket{\Psi^{(N-1)}(\lambda)}
  ,
\end{eqnarray}
where $\lambda'\equiv\Theta^{(N-1)}(\lambda)$.
Accordingly, 
$\ket{n_N(\lambda', p_N)}\otimes\ket{\Psi^{(N-1)}(\lambda)}$
and $\theta(n_N; \lambda', p_N)$ with $n_N\in\set{0,1}$
are an eigenvector and the
corresponding eigenangle of $\hat{U}^{(N)}(\lambda)$, respectively.
This implies the recursion relations
\begin{eqnarray}
  \label{eq:EigAngleRecursion}
  \theta^{(N)}(n_{N}, \dots, n_1;\lambda)
  = \theta(n_{N}; \theta^{(N-1)}(n_{N-1}, \dots, n_1;\lambda), p_{N})
\end{eqnarray}
and
\begin{eqnarray}
  \label{eq:EigVectorRecursion}
  \ket{n_{N},\dots, n_1;\lambda}
  = \ket{n_{N}(\theta^{(N-1)}(n_{N-1}, \dots, n_1;\lambda))}
  \otimes\ket{n_{N-1},\dots, n_1;\lambda}
  ,
\end{eqnarray}
for $N>1$. 

Assume that the solution of \eqref{eq:EigAngleRecursion}
takes the following form:
\begin{eqnarray}
  \label{eq:theta_by_mN}
  \theta^{(N)}(n_{N}, \dots, n_1;\lambda)
  = \frac{2\pi}{2^N}\left(m_N(n_{N}, \dots, n_1) 
    + d_N\frac{\lambda}{2\pi}\right)
  .
\end{eqnarray}
A normalized slope $d_N$ of the eigenangle and
a ``principal quantum number'' $m_N(n_{N}, \dots, n_1)$ are
defined
below.
For the sake of brevity, we will sometimes omit the arguments 
$(n_{N}, \dots, n_1)$ of $m_N$, 
so that
$\theta^{(N)}(n_{N}, \dots, n_1;\lambda)$
is written 
as $\theta^{(N)}(m_N;\lambda)$.
For $N=1$, we have $m_1(n_1) = n_1$ and $d_1 = p_1$.
For $N>1$,
\eqref{eq:EigAngleRecursion} 
provides
the 
recursion relations: 
\begin{equation}
  \label{eq:recusion_mN}
  m_N
  = 2^{N-1}n_N + p_N m_{N-1}
  \qquad\mbox{and}\qquad
  d_N 
  = p_N d_{N-1}
  .
\end{equation}
Hence we have
\begin{eqnarray}
  \label{eq:mN}
  m_N
  &
  = 2^{N-1} n_N 
  + \sum_{j=1}^{N-1} \left(\prod_{k=j+1}^N p_k\right) 2^{j-1} n_j
  ,\\
  \label{eq:dN}
  d_N 
  &
  = \prod_{k=1}^N p_k
  .
\end{eqnarray}
The solution of the recursion relation~\eqref{eq:EigVectorRecursion}
of the eigenvectors 
is
\begin{eqnarray}
  &
  \ket{n_{N},\dots, n_1;\lambda}
  = 
  \bigotimes_{j=1}^{N}\ket{n_j(\theta^{(j-1)}(m_{j-1};\lambda), p_j)}
  .
\end{eqnarray}

We remark on the spectral degeneracy of $\hat{U}^{(N)}(\lambda)$.
From \eqref{eq:theta_by_mN} and~\eqref{eq:mN}, 
the spectral degeneracy is absent if and only if 
all 
$p_2, \dots, p_N$
are odd.
Note that the degeneracy condition is independent of $p_1$ and $\lambda$.
Because we focus on the case that there is no spectral degeneracy 
along the path $C$, we will assume that $p_2, \dots, p_N$ are odd 
in the following.

\section{Eigenangle anholonomy}
\label{sec:eigenangle}

We here examine the eigenangle anholonomy 
of the family of quantum circuits described by $\hat{U}^{(N)}(\lambda)$
\eqref{eq:defUN}. 
To achieve this, we keep track of the parametric dependence of
the eigenangle $\theta^{(N)}(n_N,\dots, n_1;\lambda)$ 
along the cycle $C$, where $\lambda$ is increased by 
a period $2\pi$.
The parametric change of the eigenangle 
can be expressed as
\begin{eqnarray}
  \fl
  \label{eq:srBalance}
  \theta^{(N)}(n_N,\dots, n_1;\lambda+2\pi)
  = \theta^{(N)}(s^{(N)}(n_N,\dots, n_1; C);\lambda)
  + 2\pi r^{(N)}(n_N,\dots, n_1; C)
  ,
\end{eqnarray}
where 
$s^{(N)}(n_N,\dots,n_1;C)$
is a collection of $N$ quantum numbers
and $r^{(N)}(n_N,\dots,n_1;C)$ is an integer.
In this section, we explain how $s^{(N)}(n_N,\dots,n_1;C)$ and
$r^{(N)}(n_N,\dots,n_1;C)$ are recursively determined.
They completely characterize 
not only the eigenangle anholonomy but also the eigenspace anholonomy 
when the eigenangles are not degenerate. 
The reason is that eigenangles have one-to-one correspondence to the 
eigenspaces if the spectral degeneracy is absent.
On the other hand,
because the explicit expressions of these quantities are
complicated in general,
we postpone 
obtaining
explicit solutions;
several examples will be shown in Section~\ref{sec:examples}.

The recursive structure in $\hat{U}^{(N)}(\lambda)$ determines
$s^{(N)}(n_N,\dots,n_1;C)$ and $r^{(N)}(n_N,\dots,n_1;C)$.
For brevity, we will omit the argument $C$ in the following.
Let $s^{(N)}_j(n_N,\dots, n_1)$ denote the quantum number 
of $j$-th qubit in $s^{(N)}(n_N,\dots, n_1)$ ($1\le j\le N$). 
We note that $s^{(N)}_j(n_N,\dots, n_1)$ is either $0$ or $1$.
For $j < N$, we obtain
\begin{eqnarray}
  s^{(N)}_j(n_N,\dots, n_1) = s^{(N-1)}_j(n_{N-1},\dots, n_1)
  ,
\end{eqnarray}
using the recursive structure in $\hat{U}^{(N)}(\lambda)$.
Hence, $s^{(N)}(n_N,\dots, n_1)$  is determined by
$s^{(j)}_j(n_j,\dots, n_1)$ ($1\le j \le N$).

We
need to find $s^{(N)}_N(n_N,\dots, n_1)$ 
and $r^{(N)}(n_N,\dots, n_1)$.
For $N=1$, we obtain, 
from the analysis of the single qubit case 
in Section~\ref{sec:minimal},
$s^{(1)}_1(n_1) = s(n_1, p_1)$ and $r^{(1)}(n_1) = r(n_1, p_1)$,
where $s(n, p)$ and $r(n, p)$ are defined 
in \eqref{eq:sr_single}.
In order to clarify the case 
of
$N>1$, we first examine 
the change 
of
the principal quantum number $m_N$ 
by an increment of $\lambda$ from $0$ to $2\pi$ using
~(\ref{eq:srBalance}) and (\ref{eq:theta_by_mN}):
\begin{eqnarray}
  \label{eq:mN_balance}
  m_N(s^{(N)}(n_{N}, \dots, n_1))
  = m_N(n_{N}, \dots, n_1) + d_N - 2^N r^{(N)}(n_{N}, \dots, n_1)
  .
\end{eqnarray}
Thus
we have
\numparts
\begin{eqnarray}
  \label{eq:sN_recur}
    s^{(N)}_N(n_{N}, \dots, n_1)
    &
    = s(n_N, p_N r^{(N-1)}(n_{N-1}, \dots, n_1))
    ,
    \\
  \label{eq:rN_recur}
    r^{(N)}(n_{N}, \dots, n_1)
    &
    = r(n_N, p_N r^{(N-1)}(n_{N-1}, \dots, n_1))
    .
\end{eqnarray}
\endnumparts
We note that it is generally 
difficult to 
write down the explicit expressions of the solution of 
\eqref{eq:sN_recur} and \eqref{eq:rN_recur}.
This
problem will be dealt with 
in Section~\ref{sec:examples}.

For the time being,
we suppose that we have 
the solutions
$s^{(N)}(n_N,\dots,n_1)$ and $r^{(N)}(n_N,\dots,n_1)$
to examine 
gauge invariants.
We also assume 
that the 
spectral degeneracies are absent, i.e., $p_2,\dots,p_N$ are odd,
as explained in the previous section.
The
permutation matrix $\pma^{(N)}$ is precisely determined by
$s^{(N)}(n_N,\dots, n_1)$:
\begin{equation}
  \label{eq:PmaN}
  \{\pma^{(N)}\}_{(n'_N,\dots, n'_1), (n_N,\dots, n_1)}
  \equiv 
  \prod_{j=1}^N \delta_{n'_j,s^{(N)}_j(n_N,\dots, n_1)}
  .
\end{equation}
Note that the 
correspondence between $(n_N,\dots, n_1)$ and
$m_N(n_N,\dots, n_1)$ are one-to-one.
When we employ the ``$m_N$''-representation, instead of 
the $(n_N,\dots, n_1)$ representation, the matrix elements
of $\pma^{(N)}$ 
have simpler forms.
Because the change induced by a cycle $C$ is 
represented by the shift of $m_N$ 
by 
$d_N (\mbox{mod\ } 2^N)$,
we have
\begin{eqnarray}
  \{\pma^{(N)}\}_{m_N',m_N}
  = \delta_{m_N',m_N+d_N\mod 2^N}
  .
\end{eqnarray}
When $d_N$ is odd, or equivalently 
when $p_1$ is odd
(see \eqref{eq:dN}), 
$m_N (\mbox{mod $2^N$})$ 
returns to the initial 
value only after $2^N$ repetitions of 
evolution along
$C$. Hence the permutation matrix 
$\pma$ describes a cycle.
On the other hand, if $d_N$ is even, 
$m_N\  ({\rm mod\ } 2^N)$ 
returns with smaller number of repetitions
and the period depends on $d_N$.

\section{Eigenspace anholonomy}
\label{sec:eigenspace}
We examine the eigenspace anholonomy for $\hat{U}^{(N)}(\lambda)$
along
the adiabatic cycle $C$.
We 
find
the holonomy matrix $M^{(N)}(C)$ of
$\hat{U}^{(N)}(\lambda)$ 
and the geometric phase $\gamma(C)$~\eqref{eq:GammaByBarC}
for the adiabatic cycle $C$ 
based upon
our gauge theoretic formalism described 
in Section~\ref{sec:formalism}.

For the sake of simplicity, we focus on the case that all 
$p_j$'s are odd 
in this 
section, which guarantees that
the spectral degeneracy is absent. 
Each
eigenvalue and eigenspace 
returns 
to their initial ones only after $2^N$ cycles
along
$C$.
In other words, this assumption implies the presence of 
the one-to-one correspondence between quantum numbers 
$(n_N,\dots,n_1)$ and $m_N(n_N,\dots,n_1)$. Hence we will often 
abbreviate $(n_N,\dots,n_1)$ as $m_N$ in the following.
Also, for the sake of brevity, we will 
drop the index 
$C$.

As shown in Section~\ref{sec:formalism},
$M^{(N)}$ can be decomposed into
a product of the permutation matrix $\pma^{(N)}$ and 
a diagonal unitary matrix, i.e., 
\begin{eqnarray}
  \label{eq:Mproduct}
  \{M^{(N)}\}_{m'_N, m_N}
  = \{\pma^{(N)}\}_{m'_N, m_N}
  \sigma^{(N)}(m_N)
  ,
\end{eqnarray}
where $\sigma^{(N)}(m_N)$ is 
a
unimodular complex number.
Because 
$\pma^{(N)}$ has been
already obtained in the previous section,
our task here is to determine the phase factor 
$\sigma^{(N)}(m_N)$.

Let us
obtain a recursion relation of $M^{(N)}$ 
(\eqref{eq:Mrecursion} below) 
using the gauge connections. 
The eigenvectors of $\hat{U}^{N}(\lambda)$ 
form a non-Abelian gauge connection 
\begin{eqnarray}
  \{A^{(N)}(\lambda)\}_{m'_N, m_N}
  &
  \equiv 
  \bra{m_N';\lambda}
  \left[\rmi\partial_{\lambda}\ket{m_N;\lambda}\right]
  ,
\end{eqnarray}
and the diagonal connection
\(
  \{A^{{\rm D}(N)}(\lambda)\}_{m'_N, m_N}
  \equiv 
  \delta_{m'_N, m_N}
  \{A^{(N)}(\lambda)\}_{m'_N, m_N}
  ,
\)
which are $2^N\times 2^N$ Hermite matrices.
We 
now
obtain a recursion relation for $A^{(N)}(\lambda)$.
For $N=1$, \eqref{eq:EigenvectorOfUp}, \eqref{eq:EigenvectorOfDown} 
and the definition of the
gauge connection imply
\begin{eqnarray}
  \label{eq:A1}
  A^{(1)}(\lambda)= \frac{2-p}{4}Y
  ,\quad\mbox{and}\quad
  A^{\mathrm{D}(1)}(\lambda)=0
  .
\end{eqnarray}
Namely, for $N=1$,
each eigenvector 
satisfies the parallel transport condition within each eigenspace.
For $N>1$, the Leibniz rule suggests a decomposition 
$
A^{(N)}(\lambda)
= A_{\rm H}^{(N)}(\lambda) + A_{\rm L}^{(N)}(\lambda)$, where
\begin{eqnarray}
  \fl
  \{A_{\rm H}^{(N)}(\lambda)\}_{m'_N,m_N}
  &
  \equiv
  \left\{A(\Theta'(\lambda), p_N)\right\}_{n_N',n_N}
  \partial_{\lambda}\Theta'(\lambda)
  \delta_{m'_{N-1},m_{N-1}}
  \\
  \fl
  \label{eq:ALrecur}
  \{A_{\rm L}^{(N)}(\lambda)\}_{m'_N,m_N}
  &
  \equiv
  \bracket{n_N'(\Theta'(\lambda), p_N)}%
  {n_N(\Theta(\lambda), p_N)}
  \left\{A^{(N-1)}(\lambda)\right\}_{m'_{N-1},m_{N-1}}
  .
\end{eqnarray}
Here
$\Theta(\lambda)\equiv\theta^{(N-1)}(m_{N-1};\lambda)$
and 
$\Theta'(\lambda)\equiv\theta^{(N-1)}(m_{N-1}';\lambda)$
are introduced for abbreviations.
We also have a similar recursion relation for ``diagonal'' gauge connection 
$A^{{\rm D}(N)}(\lambda)$.
Because we have chosen the gauge $A^{{\rm D}(N)}(\lambda)=0$ for $N=1$
(see, \eqref{eq:A1}),
it is straightforward to obtain
$A^{{\rm D}(N)}(\lambda)=0$ for 
an arbitrary
$N$ from the recursion relations.

We obtain an explicit expression of $A_{\rm H}^{(N)}(\lambda)$
from the eigenangle of $N$-qubit system~\eqref{eq:theta_by_mN}
and the gauge connection of the single-qubit system~\eqref{eq:A1}:
\begin{eqnarray}
  \label{eq:AHN}
  A_{\rm H}^{(N)}(\lambda)
  &
  = \frac{(2-p_N) d_{N-1}}{2^{N+1}}Y\otimes1^{(N-1)}
  .
\end{eqnarray}

Next we examine $A_{\rm L}^{(N)}(\lambda)$ in our model. 
From \eqref{eq:EigenvectorOfUp}, \eqref{eq:EigenvectorOfDown} 
and~\eqref{eq:ALrecur}, we obtain
\begin{eqnarray}
  A_{\rm L}^{(N)}(\lambda)
  = \frac{1+Y}{2}\otimes A_{+}^{(N-1)}(\lambda)
  + \frac{1-Y}{2}\otimes A_{-}^{(N-1)}(\lambda)
  ,
\end{eqnarray}
where
\begin{eqnarray}
  A_{\pm}^{(N)}(\lambda)
  &
  =
  \rme^{\pm\rmi \pi(2-p_{N+1})2^{-N-1} J_{\rm D}^{(N)}}
  A^{(N)}(\lambda)\;
  \rme^{\mp\rmi \pi(2-p_{N+1})2^{-N-1} J_{\rm D}^{(N)}}
  ,
\end{eqnarray}
and
\begin{eqnarray}
  \label{eq:defJD}
  \left\{J_{\rm D}^{(N)}\right\}_{m'_{N},m_N}
  \equiv
  m_{N}
  \delta_{m'_{N},m_N}
  .
\end{eqnarray}
Accordingly we have
\begin{eqnarray}
  \fl
  \label{eq:ALNrec}
  A_{\rm L}^{(N)}(\lambda)
  &
  = 
  \rme^{\rmi {\pi(2-p_{N})}{2^{-N}} Y\otimes J_{\rm D}^{(N-1)}}
  \left\{1\otimes A^{(N-1)}(\lambda)\right\}
  \rme^{-\rmi {\pi(2-p_{N})}{2^{-N}} Y\otimes J_{\rm D}^{(N-1)}}
  .
\end{eqnarray}
Now it is straightforward to prove that $A_{\mathrm{L}}^{(N)}(\lambda)$ and
$A^{(N)}(\lambda)$ are independent of $\lambda$.
First, we assume the independence of $A^{(N-1)}(\lambda)$ from $\lambda$
for $N>1$.
This assumption implies that the $A_{\mathrm{L}}^{(N)}(\lambda)$
is also independent of $\lambda$, from \eqref{eq:ALNrec}.
Because $A_{\mathrm{H}}^{(N)}(\lambda)$~\eqref{eq:AHN} 
is also independent of $\lambda$, we conclude the independence of
$A^{(N)}(\lambda)$ from $\lambda$.
Second, $A^{(1)}(\lambda)$ is independent of $\lambda$ 
(see, \eqref{eq:A1}).
Hence we conclude $A_{\mathrm{L}}^{(N)}(\lambda)$ and
$A^{(N)}(\lambda)$ for $N\ge 1$ are independent of $\lambda$.

In addition,
it is straightforward to see 
$[A_{\rm H}^{(N)}(\lambda), A_{\rm L}^{(N)}(\lambda)] = 0$.
Hence the anti-path ordered product is simplified as
\(
  M^{(N)}
  = \rme^{-\rmi2\pi A_{\rm H}^{(N)}(\lambda)}
  \rme^{-\rmi2\pi A_{\rm L}^{(N)}(\lambda)}
  .
\)
This implies a recursion relation for the holonomy matrices:
\begin{eqnarray}
  \label{eq:Mrecursion}
  \fl
  M^{(N)}
  &
  = 
  \frac{1+Y}{2}\otimes
  \left(
    \rme^{\rmi {\pi(2-p_{N})}{2^{-N}} J_{\rm D}^{(N-1)}}
    M^{(N-1)}
    \rme^{-\rmi {\pi(2-p_{N})}{2^{-N}} (J_{\rm D}^{(N-1)}+d_{N-1})}
  \right)
  \nonumber\\
  \fl
  &\quad{}
  +\frac{1-Y}{2}\otimes
  \left(
    \rme^{-\rmi {\pi(2-p_{N})}{2^{-N}} J_{\rm D}^{(N-1)}}
    M^{(N-1)}
    \rme^{\rmi {\pi(2-p_{N})}{2^{-N}} (J_{\rm D}^{(N-1)}+d_{N-1})}
  \right)
  .
\end{eqnarray}
From 
this recursion relation,
we obtain the recursion relation for the matrix element of $M^{(N)}$
\begin{eqnarray}
  \fl
  \label{eq:MNelemRecur}
  \left\{M^{(N)}\right\}_{m'_N,m_N}
  &
  = 
  \delta_{n_N', s(n_N, p_{N}r^{(N-1)})}
  (-1)^{r(n_N, p_{N}r^{(N-1)})}
  \left\{M^{(N-1)}\right\}_{m'_{N-1},m_{N-1}}
  ,
\end{eqnarray}
which is shown in \ref{sec:MNelementRecursion}.

We 
now
obtain $\sigma^{(N)}(m_N)$, which is a constituent 
of $M^{(N)}$ (see \eqref{eq:Mproduct}).
For $N=1$. it is 
easy 
to see
$\sigma^{(1)}(n_1) = (-1)^{r(n_1,p_1)}$.
From 
\eqref{eq:MNelemRecur}, 
we 
find
\begin{eqnarray}
  &
  \left\{M^{(N)}\right\}_{m'_N, m_N}
  = 
  \{\pma^{(N)}\}_{m'_N, m_N}
  (-1)^{r^{(N)}}
  \sigma^{(N-1)}(m_{N-1})
  ,
\end{eqnarray}
which implies a recursion relation for $\sigma^{(N)}$:
\begin{eqnarray}
  \label{eq:r-recur}
  \sigma^{(N)}(m_N)
  =
  (-1)^{r^{(N)}(m_N)}
  \sigma^{(N-1)}(m_{N-1})
  .
\end{eqnarray}
Hence we have
\begin{equation}
  \sigma^{(N)}(n_N,\dots,n_1)
  = (-1)^{\sum_{k=1}^N r^{(k)}(n_k,\dots,n_1)}
  ,
\end{equation}
where we set $r^{(1)}(n_1)\equiv r(n_1,p_1)$.
Note that an explicit expression of $r^{(k)}(n_k,\dots,n_1)$
depends on 
$p_j$'s.

We examine Manini-Pistolesi's gauge invariant $\gamma^{(N)}$.
From \eqref{eq:Gamma_by_sigma}, we obtain
\begin{eqnarray}
  \rme^{\rmi\gamma^{(N)}}
  =
  \prod_{n_{N}=0}^{1}\cdots\prod_{n_1=0}^{1}
  \sigma^{(N)}(n_{N},\dots, n_1)
  .
\end{eqnarray}
In particular, we have already obtained
$\rme^{\rmi\gamma^{(N)}}=-1$ for $N=1$. For $N>1$, 
\eqref{eq:r-recur} implies
\(
  \prod_{n_N=0}^1
  \sigma^{(N)}(n_N,\dots, n_1)
  = \prod_{n_N=0}^1(-1)^{r^{(N)}(n_N,\dots, n_1)}
\)
due to
$\left\{\sigma^{(N-1)}(n_{N-1},\dots, n_1)\right\}^2 = 1$.
Thus
we have
\begin{eqnarray}
  \label{eq:phaseFactorbyNu}
  \rme^{\rmi\gamma^{(N)}}
  &
  =
  (-1)^{\nuK^{(N)}}
  ,
\end{eqnarray}
where 
\begin{eqnarray}
  \label{eq:nuKN}
  \nuK^{(N)}
  \equiv
  \sum_{n_{N}=0}^{1}\cdots\sum_{n_1=0}^{1} r^{(N)}(n_N,\dots, n_1)
  .
\end{eqnarray}

Recall that
$\nuK^{(1)}_1=p_1$ for $N=1$.
From \eqref{eq:rN_recur}, we find
$\sum_{n_N=0}^{1} r^{(N)}(n_N,\dots, n_1)
= p_N r^{(N-1)}(n_{N-1},\dots, n_1)$
for $N>1$, because
$[j/2] + [(1+j)/2] = j$ holds for an arbitrary integer $j$.
It leads us to
a recursion relation
$\nuK^{(N)}
  = p_N \nuK^{(N-1)}
$, whose 
solution 
is
\begin{eqnarray}
  \label{eq:nuK_by_dN}
  \nuK^{(N)}
  &
  = d_N
  .
\end{eqnarray}
Since we consider the case that no spectral degeneracy exists implying 
$d_N$ is odd, we find
\begin{eqnarray}
  \rme^{\rmi\gamma^{(N)}}
  = -1
\end{eqnarray}
for $N>0$.

\section{Examples}
\label{sec:examples}

We examine several examples in this section.
To complete the characterization of the eigenangle and eigenspace 
anholonomies of $\hat{U}^{(N)}(\lambda)$, we need to obtain
the explicit expressions of 
$s^{(N)}(n_N,\dots,n_1)$ and $r^{(N)}(n_N,\dots,n_1)$,
as explained in Section~\ref{sec:eigenangle}.
This requires to solve
\eqref{eq:sN_recur} and \eqref{eq:rN_recur},
whose
solution precisely depends on the set of integer parameters 
$\set{p_j}_{j=1}^N$.
First, we 
consider
the simplest case $p_j=1$ for all $j$.
Because we obtain the subsequent cases through the modifications of the 
simplest 
one,
the study of the first case offers the basis of the
following analysis.
Second, we replace $p_1$ 
of
the simplest case with an even integer.
It is shown that the permutation matrix 
$\pma^{(N)}(C)$ 
of the second case 
consists of two cycles.
Third, we replace $p_J$ in the simplest case with an odd integer.
These examples indicates that there are various types of 
$\pma^{(N)}(C)$
associated with the choices of $\set{p_j}_{j=1}^N$.

\begin{figure}
  \centering
  \includegraphics[width=10cm]{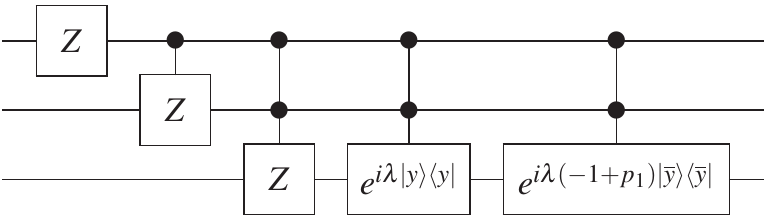}
  \caption{%
    The quantum circuit $\hat{U}^{(3)}(\lambda)$ for
    three qubits with $p_2=p_3=1$.
    $\ket{\overline{y}}$ is orthogonal to $\ket{y}$ and normalized.
    We examine the cases $p_1 = 1$ and $2$ in 
    \S~\ref{subsec:free} and \S~\ref{subsec:p1even}, respectively.
    The first examples shown in \S~\ref{subsec:single-odd}
    correspond to the case $p_1=3$.
    The convention for the controlled qubit is explained in
    the caption of figure~\ref{fig:recursion}.}
  \label{fig:3qubits}
\end{figure}

\begin{figure}
  \centering
  \includegraphics[width=10cm]{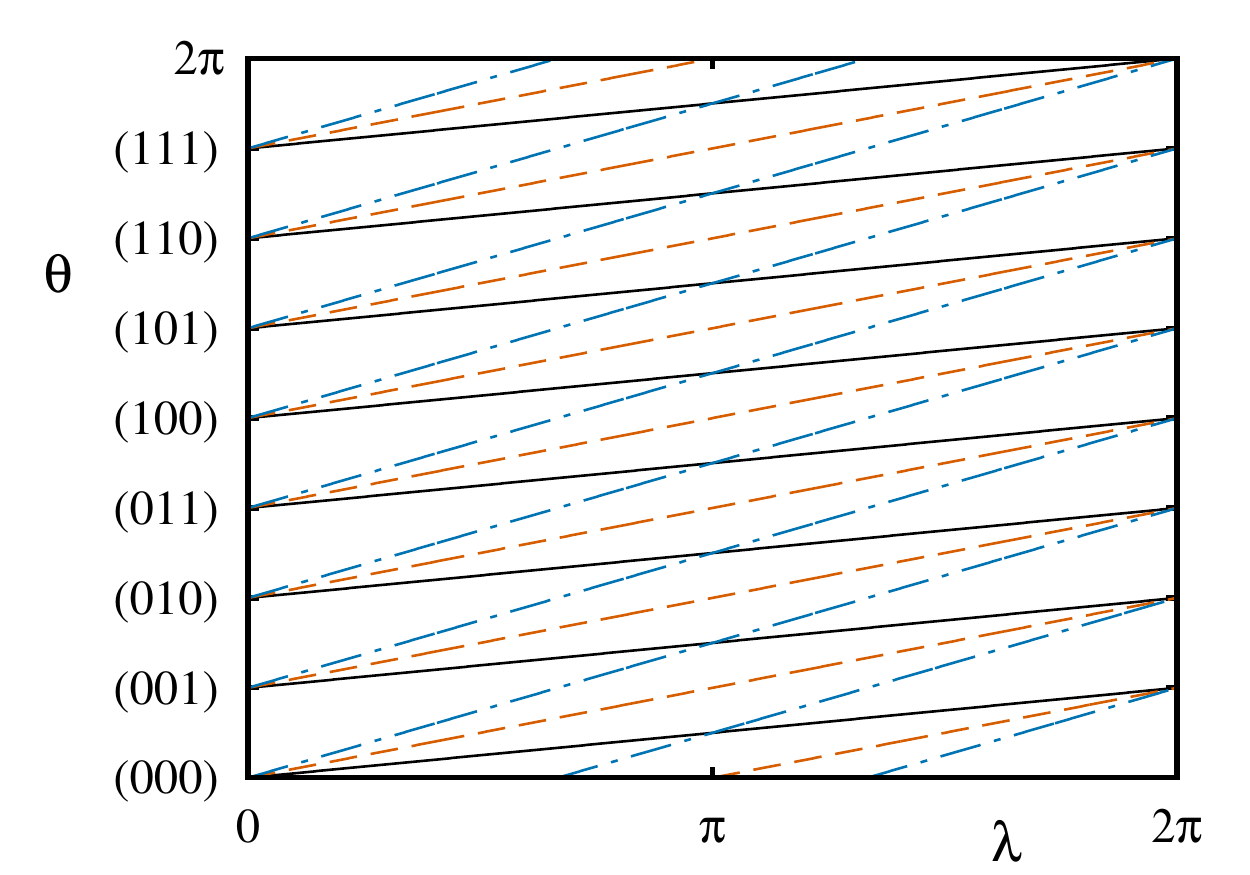}
  \caption{%
    The eigenangles $\theta^{(3)}(n_3,n_2,n_1; \lambda)$ 
    (\ref{eq:theta_by_mN})
    of $\hat{U}^{(3)}(\lambda)$
    with $p_2 = p_3 = 1$ (see figure~\ref{fig:3qubits}).
    The labels of the vertical axis indicate
    the quantum numbers $(n_3, n_2, n_1)$ of 
    the eigenstates at $\lambda=0$.
    Three families of lines correspond to
    $p_1=1$ (bold), $p_1 = 2$ (dashed) and $p_1=3$ (dash-dotted).
    Note that $p_1$ determines the slope of 
    $\theta^{(3)}(n_3,n_2,n_1; \lambda)$ .
    Hence different $p_1$ induces different itineraries of quantum numbers
    along adiabatic cycles.
    In \S~\ref{subsec:free}, we examine the simplest case $p_1=1$.
    When $p_1$ is an even integer, $\pma(C)$ is decomposed
    into multiple cycles (see (\ref{eq:p2even}) and (\ref{eq:p2odd}) 
    in \S~\ref{subsec:p1even}).
    On the other hand, odd $p_1$ ensures that $\pma(C)$
    describes a cycle (see (\ref{eq:p1_3cycle}) in \S~\ref{subsec:single-odd},
    where $p_1=3$ is assumed).
  }
  \label{fig:3qubits_levels}
\end{figure}

\subsection{%
    $p_j=1$ for all $j$: the simplest case 
}
\label{subsec:free}
We here examine the simplest case 
of
$p_j=1$ for all 
$j$
(see figures~\ref{fig:3qubits} and~\ref{fig:3qubits_levels}),
which
can be
reduced to a quantum map under rank-$1$ 
perturbation~\cite{Miyamoto-PRA-76-042115,Tanaka-EPL-96-10005}.
From \eqref{eq:mN}, we have
\begin{eqnarray}
  \label{eq:binaryExpansion_m_N}
  m_N (n_N,\dots,n_1)
  &
  = \sum_{j=1}^N 2^{j-1} n_j
  ,
\end{eqnarray}
which means
that
$n_N,\dots,n_1$ 
form
coefficients of the binary expansion of
the principal quantum number $m_N$. This provides a
direct relation
between $m_N$ and $(n_N,\dots,n_1)$.
It is straightforward to solve 
\eqref{eq:sN_recur} and \eqref{eq:rN_recur}:
\numparts
\begin{eqnarray}
  \label{eq:s_simplest}
  s^{(N)}_N(n_{N}, \dots, n_1)
  &
  =
  \cases{
    \overline{n_{N}} & \mbox{for $n_{N-1}\cdots n_1 = 1$}\\
    n_{N} & \mbox{otherwise}\\
    }
  \\
  \label{eq:r_simplest}
  r^{(N)}_N(n_{N}, \dots, n_1) 
  &
  = n_N \cdots n_1
  ,
\end{eqnarray}
\endnumparts
where we define the product of quantum numbers as
\begin{eqnarray}
  n_k \cdots n_j
  &
  =
  \cases{
    \prod_{l=j}^k n_{l}& \mbox{for $k \ge j$}\\
    1& \mbox{otherwise}\\
  }
  ,
\end{eqnarray}
throughout this paper.

The principal quantum number $m_N$ increases by one after parametric 
evolution along $C$ once due to $d_N=1$:
$
m_N = 
0\mapsto 1\mapsto \dots 
2^{N-1}-1
\mapsto 0\mapsto\dots$, whose 
period is $2^N$.
On the other hand, the itinerary of $n_N,\dots,n_1$ of $N=3$, for example,
is
\[
  000 \mapsto 
  001 \mapsto 
  010 \mapsto 
  011 \mapsto 
  100 \mapsto 
  101 \mapsto 
  110 \mapsto 
  111 \mapsto 
  000 \mapsto\dots
  .
\]

\subsection{%
    The simplest case perturbed with $p_1 \neq 1$: 
    $p_1$ is even
}
\label{subsec:p1even}
We examine the case 
of
$p_j=1$ for $j>1$ and $p_1$ is 
an
even integer, 
which corresponds to the simplest case perturbed with a certain 
even $p_1 \neq 1$
(see figures~\ref{fig:3qubits} and~\ref{fig:3qubits_levels}).
Note that the mapping between $n_N,\dots,n_1$ and $m_N$ is independent
of $p_1$,
while
the spectrum depends on $p_1$:
\begin{eqnarray}
  \theta^{(N)} (n_N,\dots,n_1)
  &
  = \frac{2\pi}{2^N}\left(m_N(n_N,\dots,n_1) + p_1\frac{\lambda}{2\pi}\right)
  .
\end{eqnarray}
Hence 
as $\lambda$ increases by every $2\pi$, 
the principal quantum number $m_N$ also increases by $p_1$.
It means that
$m_N \mod 2^N$ returns to the initial point after 
$2^{N-1}$ cycles of $C$. Furthermore, a suitable choice of $p_1$ make
the period even shorter. This is the crucial difference from the 
case that $p_1$ is 
odd, as discussed later.

For the sake of simplicity, we assume $p_1 = 2$ 
so as to find
the explicit expressions of $s^{(N)}$ and $r^{(N)}$.
For $N=1$, we have
$s^{(1)}_1(n_1) = n_1$, $r^{(1)}(n_1) = 1$.
For 
$N\ge2$, 
one finds
\numparts
\begin{eqnarray}
  s^{(N)}_N(n_{N}, \dots, n_1)
  &
  = 
  \cases{
    \overline{n_{N}} & \mbox{for $n_{N-1}\cdots n_2 = 1$}\\
    n_{N} & \mbox{otherwise}\\
  }
  ,
  \\
  r^{(N)}_N(n_{N}, \dots, n_1)
  &
  = n_N\cdots n_2
  .
\end{eqnarray}
\endnumparts
Hence the first qubit is decoupled 
from
others.
The rest of the qubits are equivalent 
to
the simplest 
case
with $N-1$ qubits.

For 
$p_1=2$, the permutation matrix 
$\pma^{(N)}(C)$ 
forms
two 
sorts of
cycles 
consisting of 
even 
or
odd 
$m_N$'s.
For $N=3$, the case of even $m_N$ reads
\begin{eqnarray}
  \label{eq:p2even}
  000 \mapsto 
  010 \mapsto 
  100 \mapsto 
  110 \mapsto 
  000 \mapsto 
  \dots
  ,
\end{eqnarray}
while for odd $m_N$'s 
\begin{eqnarray}
  \label{eq:p2odd}
  001 \mapsto 
  011 \mapsto 
  101 \mapsto 
  111 \mapsto 
  001 \mapsto 
  \dots
  .
\end{eqnarray}
We remark that 
$n_1$ remains unchanged along these itineraries.
Namely,
the 
initial qubit is repeatedly recovered during the evolution.

\subsection{%
    The simplest case perturbed with 
      odd $p_J$
}
\label{subsec:single-odd}
Here we consider the case that 
$p_j = p\delta_{jJ} + (1-\delta_{jJ})$,
where $J$ is an integer ($0 < J \le N$) 
and $p=1+2^K$ ($K$ is a positive integer).
In other words, we obtain this model by replacing $p_J$ of the simplest
case, where we set $p_j=1$ for all $j$, with $p$.
We examine how such a tiny change affect the eigenangle and eigenspace
anholonomies.
When 
this
new model consists of smaller number qubits, i.e.,  $N<J$, 
it becomes
equivalent 
to 
the simplest model.
The effect of the replacement of $p_J$ appears only when the system
size is large, i.e., $N\ge J$.

First, we 
note that
\begin{eqnarray}
  \label{eq:C_m}
  m_N(n_N,\dots, n_1)
  = \sum_{j=1}^{N} 2^{j-1}n_j + \sum_{j=1}^{J-1}2^{K+j-1}n_j
  ,
\end{eqnarray}
where the second term in the rhs is defined as $0$ if $J=1$.
We 
also
note that the spectral degeneracy is absent
since $p$ is odd

We show explicit expressions of $s_N^{(N)}(n_{N}, \dots, n_1)$ and 
$r^{(N)}(n_{N}, \dots, n_1)$, which
govern the anholonomy in the quantum number.
Because this model agrees with the simplest case for $N<J$,
$s^{(N)}_N(n_{N}, \dots, n_1)$ and 
$r^{(N)}_N(n_{N}, \dots, n_1)$ satisfy
\eqref{eq:s_simplest} and \eqref{eq:r_simplest}
for $N<J$.
The effect of the $p_J\ne 1$, which we may call 
an ``impurity'' of the simplest model, sets in at $N=J$.
For $J \le N < J + K$, 
one finds
\numparts
\begin{eqnarray}
    s^{(N)}_N(n_{N}, \dots, n_1)
    &
    = 
    \cases{
      \overline{n_{N}} & \mbox{for $n_{N-1}\cdots n_1 = 1$}\\
      n_{N} & \mbox{otherwise}\\
    }
    ,
    \\
    r^{(N)}(n_{N}, \dots, n_1)
    &
    = (n_N\cdots n_J + 2^{K-(N-J)-1})
    n_{J-1}\cdots n_1
    .
\end{eqnarray}
\endnumparts
Hence the effect of the impurity $p_J$ on $r^{(N)}$ is largest at 
$N=J$, and 
becomes
smaller as $N$ increases for $J\le N < J+K$. 
In other words, $r^{(N)}$'s of $K$ qubits are 
directly
affected
by the impurity.
The $J+K$-th qubit is in another regime:
\begin{eqnarray}
\fl
  s^{(J+K)}_{J+K}(n_{J+K}, \dots, n_1)
  = 
  \cases{
    \overline{n_{J+K}} & 
    \mbox{for $n_{J+K-1}\cdots n_J=0$ and $n_{J-1}\dots n_1 = 1$}
    \\
    n_{J+K} & \mbox{otherwise}
  }
  .
\end{eqnarray}
The influence of the qubit with index $j$ ($j=J,\dots,J+K$)
on the qubits with larger index can be
described through a quantum number 
\begin{eqnarray}
  \label{eq:t_def}
  t^{(J,K)}(n_{J+K},\dots,n_J)
  \equiv n_{J+K} + \overline{n_{J+K}}n_{J+K-1}\cdots n_J
  ,
\end{eqnarray}
which is either $0$ or $1$. 
It seems that
the impurity 
introduces 
an effective
$(K+1)$-body interaction among
the quantum numbers $(n_{J+K},\dots,n_J)$. 
This determines $r^{(N)}$ for $N\ge J+K$ as
\begin{eqnarray}
  r^{(N)}(n_{N}, \dots, n_1)
  = 
  n_N\dots n_{J+K+1} t^{(J,K)}n_{J-1}\cdots n_1
  ,
\end{eqnarray}
where we omit the argument of $t^{(J,K)}(n_{J+K},\dots,n_J)$.
On the other hand, for $N>J+K$, we have 
\begin{eqnarray}
  \fl
  s^{(N)}_N(n_{N}, \dots, n_1)
  &
  = 
  \cases{%
    \overline{n_{N}} & 
    \mbox{for $n_{N-1}\cdots n_{J+K+1} t^{(J,K)}n_{J-1}\cdots n_1
      = 1$}
    \\
    n_{N} & \mbox{otherwise}
  }
  .
\end{eqnarray}

We consider two typical cases. 
The first example is the case of
$J=1$ and $K=1$
(see figures~\ref{fig:3qubits} and~\ref{fig:3qubits_levels}),
where the principal quantum number $m_N$ is unaffected by the 
impurity 
due to
$J=1$
(see, \eqref{eq:C_m}).
As shown above, a cycle $C$ corresponds to the increment of $m_N$
by $d_N = p_1 = 3$. 
For example, with $N=3$, we have
\begin{equation}
  0 \mapsto 
  3 \mapsto 
  6 \mapsto 
  1 \mapsto 
  4 \mapsto 
  7 \mapsto 
  2 \mapsto 
  5 \mapsto 
  0 \mapsto \dots
  .
\end{equation}
In terms of $n_j$s, the itinerary is 
\begin{equation}
  \label{eq:p1_3cycle}
  000 \mapsto 
  011 \mapsto 
  110 \mapsto 
  001 \mapsto 
  100 \mapsto 
  111 \mapsto 
  010 \mapsto 
  101 \mapsto 
  000 \mapsto \dots
  .
\end{equation}
This itinerary can be understood through the simple correspondence
between $n_N,\dots,n_1$ and $m_N$, as shown in \eqref{eq:C_m}. 
Namely, $n_N,\dots,n_1$ are coefficients of the binary expansion of $m_N$.

\begin{figure}
  \centering
  \includegraphics[width=10cm]{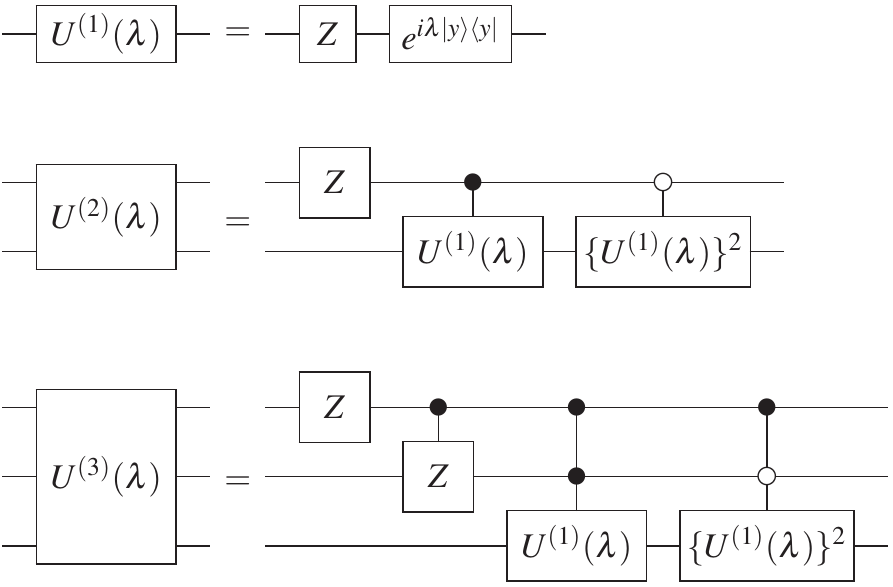}
  \caption{%
    The quantum circuits $\hat{U}^{(N)}(\lambda)$ ($N=1,2$, and $3$) 
    with $p_1=p_3=1$ and $p_2=3$.
    The cases of $N = 2$ and $3$ are rather complicated
    due to the recursive structure with $p_2\ne 1$.
    This is the second example shown in \S~\ref{subsec:single-odd}.
    The convention for the controlled qubit is explained in
    the caption of figure~\ref{fig:recursion}.}
  \label{fig:123qubits}
\end{figure}

\begin{figure}
  \centering
  \includegraphics[width=10cm]{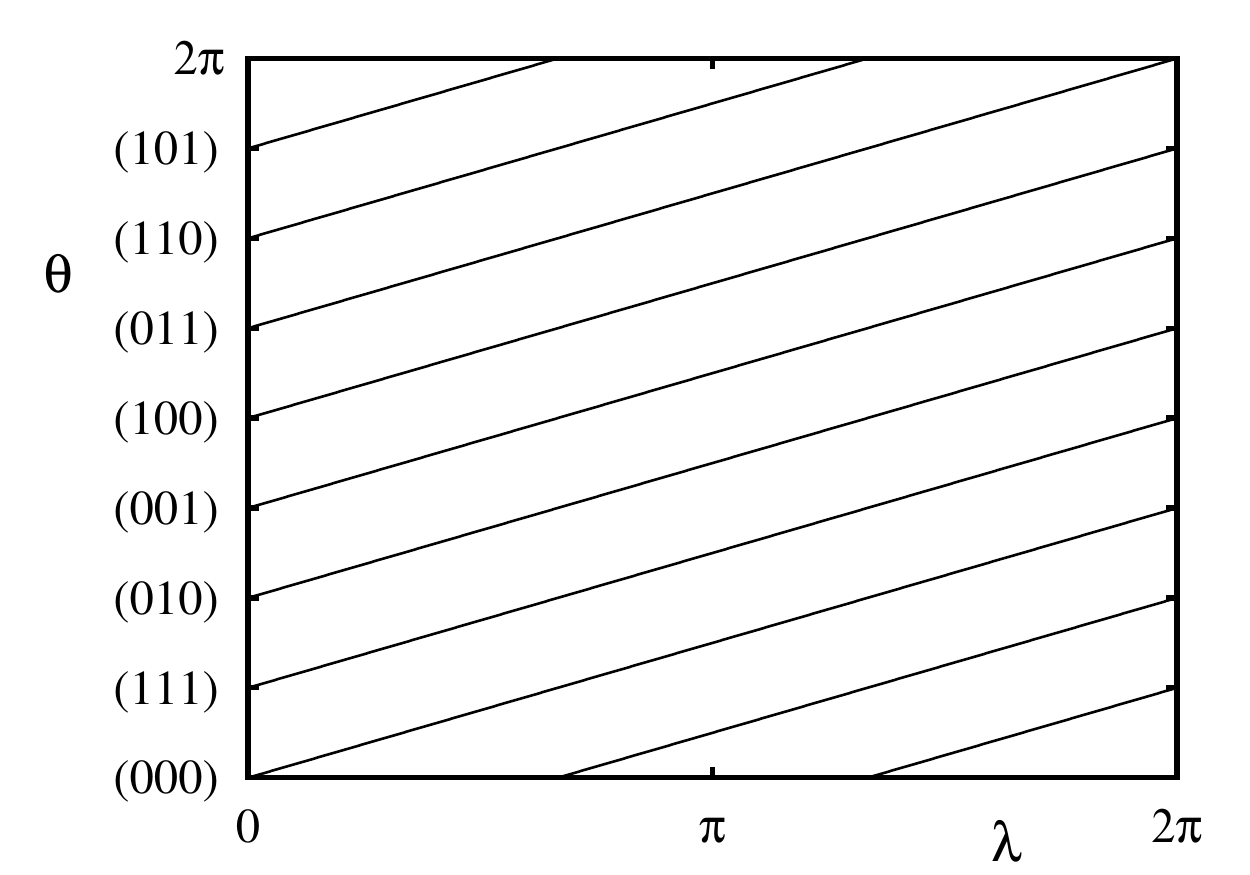}
  \caption{%
    The eigenangles $\theta^{(3)}(n_3,n_2,n_1; \lambda)$ 
    (\ref{eq:theta_by_mN}) of $\hat{U}^{(3)}(\lambda)$
    depicted in figure~\ref{fig:123qubits}.
    The labels of the vertical axis indicate
    the quantum numbers $(n_3, n_2, n_1)$ of 
    the eigenstates at $\lambda=0$.
    The order of the quantum numbers at the initial point $\lambda=0$ is 
    different from that of figure~\ref{fig:3qubits_levels}.
    Namely the initial quantum numbers are scrambled by $p_2\ne 1$.
    The corresponding itinerary of quantum numbers is shown in 
    \eqref{eq:Ccycle}.
  }
  \label{fig:123qubits_levels}
\end{figure}

The second example is the case of 
$J=2$ 
and
$K=1$
(see figure~\ref{fig:123qubits} and~\ref{fig:123qubits_levels}),
where
the effect of $p_J\ne 1$ appears not only in the eigenangle but
also in $m_N(n_N,\dots,n_1)$ (see, \eqref{eq:C_m}). 
On the other hand, the itinerary of $m_N$ induced by the repetition of
the cycle $C$ remains 
intact to be equivalent to
the case $J=1$ and $K=1$ 
discussed above;
i.e., a cycle $C$ corresponds to the increment of $m_N$ by $3$.
We depict the itinerary of quantum numbers $n_j$
induced by the cycle $C$.
For
$N=3$, we have 
$m_3 = 4 n_3 + 2 n_2 + 3 n_1$ and
\begin{equation}
  \label{eq:Ccycle}
  000 \mapsto 
  001 \mapsto 
  110 \mapsto 
  111 \mapsto 
  100 \mapsto 
  101 \mapsto 
  010 \mapsto 
  011 \mapsto 
  000\mapsto \dots
  .
\end{equation}

\section{%
  Complexity of the anholonomy
}
\label{sec:subset-sum}

The 
repetition
of the adiabatic cycle $C$ in $\hat{U}^{(N)}(\lambda)$
generates
an itinerary of the quantum numbers $(n_N,\dots,n_1)$.
We point out that the itinerary involves an NP-complete problem
in this section.

The examples in the previous sections tells us that the expressions of
$s^{(N)}_N$'s and $r^{(N)}$'s are rather simple 
when $p_j=1$ for all $j$.
In contrast, when we make 
any $p_j$
different from $1$,
the expressions of $s^{(N)}_N$'s and $r^{(N)}$'s become slightly complicated.
One may expect that 
the more $p_j$'s differ from 1, the more complicated they become, 
so does the itinerary of $(n_N,\dots,n_1)$ induced by the adiabatic 
cycle $C$.
However, 
this is not true because the itinerary of $m_N$ remains simple 
irrespective of variation of $p_j$'s; $m_N$ increases only by 
$d_N$ for a cycle $C$.
How do we justify 
this?
We will show that the cost to obtain $n_N\dots n_1$ for a given
$m_N$ is generally equivalent to 
that of finding
the solution of the subset-sum problem,
which is a NP-complete problem. We also explain 
the relationship 
between their
equivalence and the adiabatic quantum 
computation~\cite{Farhi-quant-ph-0001106} along 
an eigenangle~\cite{Tanaka-PRA-81-022320}.
For the sake of simplicity,
we focus on the case that all $p_j$ is odd. 
Then, 
the spectral degeneracy is
absent, 
and the shortest period of the itinerary 
is
$2^N$ in the unit of 
the cycle $C$.

First, we explain our task to find $(n_N\dots n_1)$ for a given 
$m_N$.
From \eqref{eq:mN}, this is to solve the following equation
\begin{eqnarray}
  \label{eq:m_vs_n}
  m_N
  &
  = \sum_{j=1}^N w_j n_j
  ,
\end{eqnarray}
where 
\begin{eqnarray}
  w_j \equiv \left(\prod_{k=j+1}^N p_k\right) 2^{j-1}
\end{eqnarray}
for $1\le j < N$, and $w_N\equiv 2^{N-1}$.

Second, we explain the subset-sum problem~\cite{SubsetSum}.
A subset-sum problem has two 
parameters
$S$ and $t$.
$S$ is a finite set of positive integers.
$t$ is a positive integer, and is called a target. 
The
problem is to show whether there exists a set $S'\subseteq S$ 
such that $t = \sum_{s\in S'} s$.
For our purpose, we need to obtain $S'$.

Now we explain the equivalence between our problem and the subset-sum.
In our case, let $S$ be $\set{w_j}_{j=1}^N$, which is a set of positive
integers.
Finding $(n_N\dots n_1)$ that satisfy \eqref{eq:m_vs_n}
is essentially equivalent 
to
the problem to find a set of 
positive integers $S'\subseteq S$ such that $m_N = \sum_{s\in S'} s$. 
This justifies the equivalence.

This explanation has a subtle point. During a course of the 
repetition of the cycle $C$, $m_N$ can be determined only up to modulo
$2^N$. However, this does not 
break
the equivalence in the difficulty
of the solving the problems

Based upon 
this equivalence,
we examine 
whether the evolution along the adiabatic increment of $\lambda$
in $\hat{U}^{(N)}(\lambda)$ offers an efficient way to solve the
subset-sum problem. 
It
turns out that the adiabatic
approach is inefficient.
First, when $\lambda$ is varied slow enough, 
we can prepare $m_N$ to be an arbitrary positive integer.
Next, appropriate measurements of qubits ascertain $n_N,\dots, n_1$.
To make the nonadiabatic error small enough, there is a lower bound 
of the running time for quantum 
evolution. This is essentially
determined by the gap $\Delta$ between the eigenangles. 
From the exact expression of the eigenangle~\eqref{eq:theta_by_mN},
we obtain 
$
  \Delta \propto 2^{-N}
$
for large $N$. Hence the lower bound of
the running time must be exponentially large, i.e.,
the adiabatic approach is inefficient.

We remark on the inefficiency to obtain $\hat{U}^{(N)}(\lambda)$
from quantum gates. Our construction described in \eqref{eq:defUN},
of $\hat{U}^{(N)}(\lambda)$ is generally inefficient, because 
\eqref{eq:defUN} 
requires exponentially many quantum 
gates
for general $\set{p_j}_{j=0}^{\infty}$.
On the other hand, if
we impose $p_j=1$ for most $j$ except for infinitely
many $j$, the number of quantum gates required to construct
$\hat{U}^{(N)}(\lambda)$ can be a polynomial of $N$. However,
we are not certain whether such a $\hat{U}^{(N)}(\lambda)$ 
involves any NP-complete problem.

\section{Discussion}
\label{sec:discussion}

We here discuss the relationship 
between our result and
the recent work
on the topological characterization of periodically driven
systems~\cite{Kitagawa-PRB-82-235114}. 
The work in \cite{Kitagawa-PRB-82-235114}
offers
an integer 
\begin{eqnarray}
  \label{eq:nuK}
  \nuK (C)
  \equiv \frac{1}{2\pi\rmi}\oint_C \Tr[\{\hat{U}(\lambda)\}^{-1}
  \partial_{\lambda}\hat{U}(\lambda)]\rmd\lambda
  ,
\end{eqnarray}
for a family of Floquet
operator $\hat{U}(\lambda)$ in a closed path $C$ in the parameter
space of $\lambda$. 
Because $\nu(C)$ can be regarded as a gauge invariant 
for the adiabatic cycle $C$, it is worth 
comparing
it with
our gauge invariants $\pma(C)$ and $\gamma(C)$.

First, we show that $\nuK (C)$ is indeed an integer.
Let $\hat{U}(\lambda)$ denote a Floquet operator of a periodically
driven system. We assume that $\hat{U}(\lambda)$ is periodic 
in $\lambda$ and its shortest period is 
$2\pi$, which guarantees its spectrum 
$\set{\rme^{\rmi\theta_n(\lambda)}}_{n=0}^{N-1}$ is also periodic in $\lambda$,
where $N$ is the dimension of the Hilbert space and
$\theta_n(\lambda)$ is an eigenangle.
Such a
spectral periodicity implies
\begin{eqnarray}
  \label{eq:theta_vs_rn}
  \theta_{s(n)}(0) = \theta_n(2\pi) + 2\pi r(n)
  ,
\end{eqnarray}
where $s(n)$ describes a permutation over quantum numbers
$\set{0,\dots,N-1}$, and $r(n)$ is an integer.
The latter integer has an integral expression
\begin{eqnarray}
  2\pi r(n) = \int_0^{2\pi} \pdfrac{\theta_n(\lambda)}{\lambda} \rmd\lambda
  .
\end{eqnarray}
The derivative of eigenangle 
is
$
\pddiv{\theta_n(\lambda)}{\lambda}
= -\rmi\bra{n(\lambda)}\{\hat{U}(\lambda)\}^{-1} 
\{\partial_{\lambda}\hat{U}(\lambda)\}\ket{n(\lambda)},
$
where $\ket{n(\lambda)}$ is a normalized eigenvector of
$\hat{U}(\lambda)$ corresponding to 
the eigenangle $\theta_n(\lambda)$~\cite{Nakamura-PRL-57-1661}.
Hence we obtain
\begin{eqnarray}
  2\pi r(n) = -\rmi\int_0^{2\pi}
  \bra{n(\lambda)}\{\hat{U}(\lambda)\}^{-1} 
  \{\partial_{\lambda}\hat{U}(\lambda)\}\ket{n(\lambda)}
  \rmd\lambda
  .
\end{eqnarray}
A sum rule on $2\pi r(n)$ has a representation-independent expression:
\begin{eqnarray}
  \label{eq:2pir_sumrule}
  2\pi \sum_{n=0}^{N-1}r(n)
  = \frac{1}{\rmi}\int_0^{2\pi}\Tr\left[\{\hat{U}(\lambda)\}^{-1} 
    \{\partial_{\lambda}\hat{U}(\lambda)\}\right]
  \rmd\lambda
  .
\end{eqnarray}
We remark that 
this sum rule
is the key to prove
the presence of the eigenvalue and eigenspace anholonomies
in the quantum map under 
a rank-$1$ perturbation~\cite{Tanaka-PRL-98-160407,Miyamoto-PRA-76-042115}.
\eqref{eq:nuK} and~\eqref{eq:2pir_sumrule} imply
\begin{eqnarray}
  \label{eq:nuKdef}
  \nuK(C)
  \equiv \sum_{n=0}^{N-1}r(n)
  ,
\end{eqnarray}
which is an integer.
This argument suggests that $\nuK(C)$ characterizes the winding of
eigenangles, or equivalently, quasienergies.

We now compare $\nuK(C)$ with the permutation matrix $\pma(C)$.
We examine the classification of a closed path $C$ for a
family of the quantum 
circuit
on a qubit, using $\nuK(C)$.
We 
recall
that $\pma(C)$ is shown to classify $C$ into 
two classes in Sec.~\ref{sec:minimal}.
This is obtained by an inspection of $C_p$,
which 
corresponds
to the family of unitary $\hat{u}(\lambda,p)$
\eqref{eq:defU} for $0\le \lambda\le 2\pi$.
In particular, even 
and odd $p$ correspond to 
\eqref{eq:pma2even} and \eqref{eq:pma2odd}, respectively.
As for $\nuK(C)$,
it is straightforward to see $\nuK(C_p)=p$.
The stability of the topological quantity $\nuK(C)$ against 
a small deformation of closed path from $C_p$ implies that
$\nuK(C)$ of a closed path $C$ takes an arbitrary integer.
Hence $\nuK(C)$ classifies the closed paths into an infinitely
many classes.
Thus
$\nuK(C)$ 
exhibits
a detailed structure than $\pma(C)$ 
as 
for the closed paths for the space of single-qubit quantum circuits.

This conclusion does not hold for the quantum circuits with multiple
qubits.
To see this, we examine 
the quantum circuits $\hat{U}^{(N)}(\lambda)$ introduced 
in Sec.~\ref{sec:model}.
We have already obtained the corresponding integer $\nuK^{(N)}(C)$
in \eqref{eq:nuK_by_dN} to evaluate the geometric phase 
$\gamma^{(N)}(C)$. The examples shown in Sec.~\ref{subsec:single-odd}
tells us that $\pma^{(N)}(C)$ can take various 
values
even 
for a certain given
$\nuK^{(N)}(C)$. 
Thus we conclude that the role of $\pma^{(N)}(C)$ and $\nuK^{(N)}(C)$ 
are independent 
in classifying
the families of quantum circuits.

In order to provide another comparison of $\pma(C)$ and $\nuK(C)$,
we show an example in which $C$ contains the crossing of 
eigenvalues.
Note that we have 
excluded such cases so far in this paper.
We will show that $\pma(C)$ is sensitive to the eigenvalue crossing,
while $\nuK(C)$ is not~\cite{Kitagawa-PRB-82-235114}.
Let us examine the following quantum circuit 
\begin{eqnarray}
  \label{eq:defUY}
  \hat{u}_{Y}(\lambda, p)
  \equiv
  \rme^{\rmi(p-1)\lambda(1-\ketbra{y}{y})
    +\rmi\lambda\ketbra{y}{y}}\hat{Y}
  ,
\end{eqnarray}
which is obtained by replacing $\hat{Z}$ 
with
$\hat{Y}\equiv 1-2\ketbra{y}{y}$
in $\hat{u}(\lambda, p)$ \eqref{eq:defU}.
These quantum circuits are periodic in $\lambda$.
Let $C_{{Y}}$ denote the closed path in quantum circuits
specified by $\hat{u}_{Y}(\lambda, p)$ with $0\le \lambda\le 2\pi$.
The crucial point is that $C_{{Y}}$ involves 
an eigenvalue crossing, because the eigenvalues of
$\hat{u}_{Y}(\lambda, p)$ degenerate at $\lambda=0$.
Also, the eigenvector of $\hat{u}_{Y}(\lambda, p)$ is independent
of $\lambda$. Hence $\pma(C_{{Y}})$
is the identity matrix~\eqref{eq:pma2even}.
We explain that $\pma(C)$ in the vicinity of $C=C_{Y}$ is
generically different from the identity matrix.
Let us introduce another unitary matrix $\hat{Y}'$, which satisfies
$[\hat{Y}',\hat{Y}]\ne 0$.
When we replace 
$\hat{Y}$ in $\hat{u}_{Y}(\lambda, p)$ with $\hat{Y'}$, 
the corresponding family of quantum circuit
exhibits the eigenspace and eigenangle anholonomies,
according to the analysis in \cite{Tanaka-PRL-98-160407}.
Hence $\pma(C)$ is the permutation matrix whose cycle is $2$ 
~\eqref{eq:pma2odd}.
This conclusion holds even when the difference between
$\hat{Y}$ and $\hat{Y}'$ is arbitrary small.
Thus, $\pma(C)$ is sensitive to $C$ in the vicinity 
of $C_{Y}$.
In contrast, $\nu(C)=p$ is independent of $\hat{Y}$ in 
$\hat{u}_{Y}(\lambda, p)$, as long as $\hat{Y}$ is unitary.
In this sense, $\nu(C)$ is stable against the choice of $C$, even
when $C$ involves the crossing of eigenvalues.

We close this discussion with a comment on the gauge invariants
obtained here. All gauge invariants are determined by the integers
$s(n; C)$ and $r(n; C)$, which are defined in~\eqref{eq:srBalance} (see also,
\eqref{eq:theta_vs_rn}), as for the adiabatic cycle $C$ of quantum circuits
$\hat{U}^{(N)}(\lambda)$~\eqref{eq:defUN}.  The permutation matrix
$\mathfrak{S}^{(N)}$ contains all $s(n)$'s.  On the other hand,
$\nu(C)$~\eqref{eq:nuKdef} is the whole sum of $r(n)$'s.  We find that
$\nu(C)$ determines the geometric phase $\gamma(C)$ through
\eqref{eq:phaseFactorbyNu}. 
It remains to be clarified whether the intimate relationship
between $\nu(C)$ and $\gamma(C)$ holds in general.  Also, it is worth
to clarify whether other combinations of $r(n)$'s give us any useful
insights on the anholonomies.

\section{Summary}
\label{sec:summary}

In this paper, we 
have identified
the gauge invariants
$\pma(C)$ and $\gamma(C)$ that are associated with the eigenvalue 
and eigenspace anholonomies for the closed path $C$ of a family 
of unitary operators.
The unified theory of quantum anholonomy has been 
revisited
to clarify that $\gamma(C)$ is the Berry phase for the $N$-repetition 
of closed path $C^N$, where $N$ is the dimension of the relevant
Hilbert space.
By using
a family of quantum circuits that are recursively 
constructed,
these gauge invariants have been analyzed in detail.
It 
has been shown 
that a generic family of the quantum circuits 
is associated with
an NP-complete problem.
The relationship between our gauge invariants
and Kitagawa {\it et al.}'s topological integer $\nuK(C)$ 
has been also discussed.

\ack
AT wishes to thank Takuya Kitagawa for a useful conversation.
This work has been partially supported by the Grant-in-Aid for Scientific 
Research of MEXT, Japan (Grant numbers 22540396 and 21540402).
SWK was supported by 
the NRF grant funded by the Korea government (MEST) (No.~2010-0024644).

\appendix

\section{A proof of \eqref{eq:MNelemRecur}}
\label{sec:MNelementRecursion}
We will obtain 
the recursion relation~\eqref{eq:MNelemRecur} of 
the matrix element of $M^{(N)}$ from 
the recursion relation~\eqref{eq:Mrecursion} of $M^{(N)}$.

First, 
to simplify \eqref{eq:Mrecursion}, we need to evaluate
$\rme^{\rmi\alpha J_{\rm D}^{(N)}} M^{(N)} \rme^{-\rmi\alpha J_{\rm D}^{(N)}}$.
Because $J_{\rm D}^{(N)}$~\eqref{eq:defJD} is diagonal
and $M^{(N)}$ is the product of a permutation and a diagonal unitary 
matrices (see, ~\eqref{eq:Mproduct}), we have
\begin{eqnarray}
  \fl
  \left\{[J_{\rm D}^{(N)}, M^{(N)}]\right\}_{m'_N, m_N}
  = 
  \{M^{(N)}\}_{m'_N, m_N}
  \left[
    m_N(s^{(N)}(n_N,\dots, n_1))
    - 
    m_N(n_N,\dots, n_1)\right]
  .
\end{eqnarray}
Furthermore, using the recursion relation~\eqref{eq:mN_balance}
for $m_N$,
we obtain
\begin{eqnarray}
  \left\{[J_{\rm D}^{(N)}, M^{(N)}]\right\}_{m'_N, m_N}
  = 
  \{M^{(N)}\}_{m'_N, m_N}\left(d_N - 2^N r^{(N)}\right)
  ,
\end{eqnarray}
where $r^{(N)}(n_N, \dots, n_1)$ is abbreviated as $r^{(N)}$.
Now it is straightforward to show
\begin{eqnarray}
  \left\{
    \rme^{\rmi\alpha J_{\rm D}^{(N)}} M^{(N)}\rme^{-\rmi\alpha J_{\rm D}^{(N)}}
  \right\}_{m'_N, m_N}
  = 
  \left\{M^{(N)}\right\}_{m'_N, m_N}
  \rme^{\rmi\alpha
    \left(
      d_N - 2^N r^{(N)}
    \right)}
  .
\end{eqnarray}
Hence, we obtain from ~\eqref{eq:Mrecursion}
\begin{eqnarray}
  \left\{M^{(N)}\right\}_{m'_N, m_N}
  = 
  \label{eq:MNElementWithY}
  \left\{
    \rme^{-\rmi(\pi/2)(2-p_{N})r^{(N-1)}Y}
  \right\}_{n_N',n_N}
  \left(M^{(N-1)}\right)_{m'_{N-1}, m_{N-1}}
  .
\end{eqnarray}
When $r^{(N-1)}$ is even, the first factor in rhs of 
~\eqref{eq:MNElementWithY} is written as
\begin{eqnarray}
  &
  \left\{\rme^{-\rmi\pi(2-p_{N})r^{(N-1)}Y/2}\right\}_{n_N',n_N}
  = \delta_{n_N',n_N}(-1)^{-(2-p_{N})r^{(N-1)}/2}
  .
\end{eqnarray}
On the other hand, when $r^{(N-1)}$ is odd, we have
\begin{eqnarray}
  \fl
  \left\{\rme^{-\rmi\pi(2-p_{N})r^{(N-1)}Y/2}\right\}_{n_N',n_N}
  = 
  \delta_{n_N', \overline{n_N}}(-1)^{n_N}
  (-1)^{[(2-p_{N})r^{(N-1)}(n_{N-1},\dots,n_1)-1]/2}
  .
\end{eqnarray}
We 
then
obtain, using the definition of $r(n,p)$~\eqref{eq:sr_single},
\begin{eqnarray}
  &
  \left\{\rme^{-\rmi\pi(2-p_{N})r^{(N-1)}Y/2}\right\}_{n_N',n_N}
  = \delta_{n_N', s(n_N, p_{N}r^{(N-1)})}
  (-1)^{r(n_N, p_{N}r^{(N-1)})}
  .
\end{eqnarray}
Hence \eqref{eq:MNelemRecur} is 
proved.

\section*{References}

\begin{thebibliography}{10}
\expandafter\ifx\csname url\endcsname\relax
  \def\url#1{{\tt #1}}\fi
\expandafter\ifx\csname urlprefix\endcsname\relax\def\urlprefix{URL }\fi
\providecommand{\eprint}[2][]{\url{#2}}

\bibitem{Berry-PRSLA-392-45}
Berry M~V 1984 {\em Proc. R. Soc. London\/} {\bf A 392} 45

\bibitem{Cheon-PLA-248-285}
Cheon T 1998 {\em Phys. Lett. A\/} {\bf 248} 285

\bibitem{generalizedPointInteraction}
Albeverio S, Gesztesy F, Hoegh-Krohn R and Holden H
with an appendix by Exner P
2005
{\it Solvable Models in Quantum Mechanics}
2nd edn 
(Rhode Island: AMS Chelsea)
Appendix~K


\bibitem{Cheon-PLA-374-144}
Cheon T, Tanaka A and Kim S~W 2009 {\em Phys. Lett. A\/} {\bf 374} 144

\bibitem{Tanaka-PRA-82-022104}
Tanaka A and Cheon T 2010 {\it Phys. Rev. A\/} {\bf 82} 022104

\bibitem{Tanaka-PRL-98-160407}
Tanaka A and Miyamoto M 2007 {\it Phys. Rev. Lett.\/} {\bf 98}
  160407

\bibitem{Miyamoto-PRA-76-042115}
Miyamoto M and Tanaka A 2007 {\it Phys. Rev. A\/} {\bf 76} 042115

\bibitem{Tanaka-EPL-96-10005}
Tanaka A, Kim S~W and Cheon T 2011 {\it Europhys. Lett.\/} {\bf 96}
  10005

\bibitem{Wilczek-PRL-52-2111}
Wilczek F and Zee A 1984 {\it Phys. Rev. Lett.\/} {\bf 52} 2111

\bibitem{Cheon-EPL-85-20001}
Cheon T and Tanaka A 2009 {\it Europhys. Lett.\/} {\bf 85} 20001

\bibitem{Fujikawa-PRD-72-025009}
Fujikawa K 2005 {\it Phys. Rev. D\/} {\bf 72} 025009

\bibitem{Bohm-GPQS-2003}
Bohm A, Mostafazadeh A, Koizumi H, Niu Q and Zwanziger Z 2003 {\it The
  Geometric Phase in Quantum Systems\/} (Berlin: Springer)

\bibitem{Farhi-quant-ph-0001106}
Farhi E, Goldstone J, Gutmann S and Sipser M 2000 Quantum computation by
  adiabatic evolution \textit{Preprint} \eprint{quant-ph/0001106}

\bibitem{Tanaka-PRA-81-022320}
Tanaka A and Nemoto K 2010 {\it Phys. Rev. A\/} {\bf 81} 022320

\bibitem{Kitagawa-PRB-82-235114}
Kitagawa T, Berg E, Rudner M and Demler E 2010 {\it Phys. Rev. B\/} {\bf 82}
  235114

\bibitem{TANAKA-AP-85-1340}
Tanaka A and Cheon T 2009 {\it Ann. Phys. (N.Y.)\/} {\bf 324} 1340

\bibitem{Stone-PRSLA-351-141}
Stone A~J 1976 {\it Proc. R. Soc. London\/} {\bf A 351} 141

\bibitem{Samuel-PRL-60-2339}
Samuel J and Bhandari R 1988 {\it Phys. Rev. Lett.\/} {\bf 60} 2339

\bibitem{Manini-PRL-85-3067}
Manini N and Pistolesi F 2000 {\it Phys. Rev. Lett.\/} {\bf 85} 3067

\bibitem{Mead-JCP-70-2284}
Mead C and Truhlar D~G 1979 {\it J. Chem. Phys.\/} {\bf 70} 2284

\bibitem{cycle}
Georgi H 1999 {\it Lie Algebras in Particle Physics\/} 2nd edn
(Colorado: Westview press) Chap~1

\bibitem{SubsetSum}
Cormen T H {\it et al.} 2009 {\it Introduction to Algorithms} 3rd edn
(Massachusetts: The MIT Press) \S~34.5.5.

\bibitem{Nakamura-PRL-57-1661}
Nakamura K and Lakshmanan M 1986 {\it Phys. Rev. Lett.\/} {\bf 57} 1661

\end{thebibliography}

\providecommand{\newblock}{}


\end{document}